\documentclass[english,preprint]{aastex}
\usepackage{array}
\usepackage{rotfloat}
\usepackage{units}
\usepackage{graphicx}
\usepackage{amssymb}

\makeatletter

\shorttitle{MC Simulations of Globular Clusters with an IMBH}
\shortauthors{Umbreit et al.}

\makeatother

\usepackage{babel}

\begin{document}

\title{Monte Carlo Simulations of Globular Cluster Evolution. VI. \\
The Influence of an Intermediate-Mass Black Hole.}

\author{Stefan Umbreit}

\affil{Center for Interdisciplinary Exploration and Research in Astrophysics
(CIERA) \& Dept. of Physics and Astronomy, Northwestern University, 2145
Sheridan Rd, Evanston, IL 60208, USA.}

\email{<s-umbreit@northwestern.edu>}

\author{John M. Fregeau \altaffilmark{1}}

\affil{Kavli Instiute of Theoretical Physics, University of California,
Santa Barbara, CA 93106}
\altaffiltext{1}{Chandra Fellow}

\author{Sourav Chatterjee}

\affil{University of Florida, 211 Bryant Space Science Center, Florida, USA}

\and{}

\author{Frederic A. Rasio}

\affil{Center for Interdisciplinary Exploration and Research in Astrophysics
(CIERA) \& Dept. of Physics and Astronomy, Northwestern University, 2145
Sheridan Rd, Evanston, IL 60208, USA.  IL 60208}

\begin{abstract}
We present results from a series of Monte Carlo simulations investigating
the imprint of a central intermediate-mass black hole (IMBH) on the
structure of a globular cluster. We investigate the three-dimensional
and projected density profiles, and stellar disruption rates for idealized
as well as realistic cluster models, taking into account a stellar
mass spectrum and stellar evolution, and allowing for a larger, more
realistic, number of stars than was previously possible with direct
N-body methods. We compare our results to other $N$-body and Fokker-Planck
simulations published previously. We find, in general, very good agreement
for the overall cluster structure and dynamical evolution between
direct $N$-body simulations and our Monte Carlo simulations. Significant
differences exist in the number of stars that are tidally disrupted
by the IMBH, and this is most likely caused by the wandering motion
of the IMBH, not included in the Monte Carlo scheme. These differences,
however, are negligible for the final IMBH masses in realistic cluster
models, as the disruption rates are generally much lower than for single-mass
clusters. As a direct comparison to observations we construct a
detailed model for the cluster NGC 5694, which is known to possess a central
surface brightness cusp consistent with the presence of an IMBH. We find that
not only the inner slope but also the outer part of the surface brightness
profile agree well with observations. However, there is only a
slight preference for models harboring an IMBH compared to models without. 


\end{abstract}

\section{Introduction}

As recently as 10 years ago, it was generally believed that black holes (BHs)
occur in two broad mass ranges: stellar ($M_{BH}\backsimeq3-20M_{\odot}$),
produced by the core collapse of massive stars, and supermassive
($M_{BH}\sim10^{6}-10^{10}M_{\odot}$), believed to have formed in the centers
of galaxies at high redshift and grown in mass as a result of galaxy mergers
(see, e.g., Volonteri, Haardt \& Madau 2003). However, the existence of BHs
with masses intermediate between these two ranges could not be established by
observations until recently, although intermediate mass BHs (IMBHs) were
discussed by theorists more than 30 years ago (see, e.g., Wyller 1970).
Indirect evidence for IMBHs has accumulated over time from observations of
so-called ultraluminous X-ray sources (ULXs), objects with fluxes that exceed
the angle-averaged flux of a stellar mass BH accreting at the Eddington limit.
An interesting result from observations of ULXs is that many, if not most, of
them are associated with star clusters.  It has long been speculated (e.g.,
Frank \& Rees 1976) that the centers of globular clusters (GCs) may harbor BHs
with masses $\sim10^{3}{\rm {M_{\odot}}}$.  If so, these BHs affect the
distribution function of the stars, producing velocity and density cusps
(Bahcall \& Wolf 1976). While the detection of ULXs can only give indirect
evidence of the presence of IMBHs, observations of cuspy velocity profiles
would make it possible to directly determine the BH mass. However, the radius
of influence of an IMBH, defined as the radius where the orbital velocity
around the BH equals the velocity dispersion of the cluster, is very small. For
example, at a distance of $10\,{\rm kpc}$, a $10^{3}{\rm M}_{\odot}$ BH would
influence orbits within $\approx1"$ , making observations very challenging. 

Studies of the surface density profile of GCs offer a complementary
method of constraining the effects of an IMBH on the host GC stars.
A recent study by Noyola \& Gebhardt (2006) obtained central surface
brightness profiles for 38 Galactic GCs from HST WFPC2 images. They
showed that half of the GCs in their sample have slopes for the inner
surface brightness profiles that are inconsistent with simple isothermal
cores, which may be indicative of an IMBH. However, it is challenging
to explain the full range of slopes with current models. While analytical
models can only explain the steepest slopes in their sample, recent
N-body models of GCs containing IMBHs (Baumgardt et al. 2005), might
explain some of the intermediate surface brightness slopes. 

However, the disadvantage of current N-body simulations is that for
realistic cluster models, which take into account stellar evolution
and a realistic mass spectrum, the number of stars is restricted to
typically less than $\sim10^{5}$ as these simulations require a large
amount of computing time. However, many GCs are known to be very massive,
with masses reaching up to $2\times10^{6}\,{\rm {M_{\odot}}}$ resulting
in a much larger number of stars one has to deal with when modeling
these objects. In previous N-body simulations, such large-N clusters
have been scaled down to low-N systems. Scaling down can be achieved
in two ways (e.g. Baumgardt et al. 2005): either the mass of the central
IMBH $M_{BH}$ is kept constant and $N$ is decreased, effectively
decreasing the total cluster mass $M_{C}$, or the ratio $M_{BH}/M_{C}$
is kept constant, while lowering both $M_{BH}$ and $M_{C}$. As both
$M_{BH}/M_{C}$ and the ratio of $M_{BH}$ to stellar mass are important
parameters that influence the structure and dynamics of a cluster,
but cannot be held constant simultaneously when lowering $N$, it
is clear that only with the real $N$ a fully self-consistent simulation
can be achieved. Scaling becomes even more difficult once other physical
processes are included into the simulations such as stellar evolution
or stellar collisions. Using the correct number of stars in a dynamical
simulation ensures that the relative rates of different dynamical
processes (which all scale differently with $N$) are correct. 

It is clear that in order to study the evolution of old globular clusters
that might harbor IMBHs, more approximate methods have to be employed.
They fall roughly into two categories, methods that treat the cluster
as a continuum distribution function \citep{2004MNRAS.352..655A,1997PASJ...49..547T,1994MNRAS.269..241G,1991ApJ...370...60M,1978ApJ...226.1087C,1977ApJ...211..244L,1977ApJ...216..883B},
and Monte Carlo (MC) methods that use a particle based approach  \citep[see,
e.g.,][and references therein]{2007ApJ...658.1047F,2002A&A...394..345F,2001A&A...375..711F,1978ApJ...225..603S,1971Ap&SS..14..151H}.
While the former methods have been successfully
used to study the effect of a massive BH on the cluster structure
\citep{2004MNRAS.352..655A,1991ApJ...370...60M,1978ApJ...226.1087C,1977ApJ...211..244L},
the model clusters were highly idealized, consisting only of equal-mass
stars, and did not incorporate stellar evolution. Although including
additional physical processes is not impossible, it remains nevertheless
highly non-trivial for these methods. The MC method, on the other
hand, relies on a star-by-star description of the  cluster and has,
therefore, the great advantage that additional processes are easily
incorporated. 

Our group has been developing over many years a state-of-the-art MC
code, which treats many relevant processes in sufficient detail, making
direct comparison with GC observations feasible (\citealt{2007ApJ...658.1047F}
and references therein). This paper is the sixth in a series
studying the fundamental aspects of cluster dynamics using this code.
Here we will describe the changes we made to our code in order to
incorporate the effect of a massive central IMBH and carry out comparison
runs with idealized models as well as more realistic cluster simulations
published previously in the literature. In \S \ref{sec:Monte-Carlo-Method}
we briefly describe our method and the changes we made to the MC code.
We validate our code by comparing our results to previously published
results in the literature, using idealized cluster models (\S\ref{sec:Idealized-Models}),
as well as more realistic ones that include stellar evolution (\S\ref{sec:Realistic-Cluster-Models}).
In \S\ref{sub:Comparison-to-Real} we present surface brightness
profiles from our large-$N$ runs and compare them to observations
of \citet{2006AJ....132..447N}. We conclude in \S\ref{sec:Summary-and-Conclusions}.

\section{Previous Work on Globular Cluster Evolution with IMBHs\label{sec:Previous-Work-of}}

The dynamical effect of an IMBH on the surrounding stellar system
was first described by Peebles (1972), who argued that the bound stars
in the cusp around the BH must obey a shallow power-law density distribution
to account for stellar consumption near the cluster center. Analyzing
the Fokker-Planck equation in energy space for an isotropic stellar
distribution, \citet{1976ApJ...209..214B} obtained a density profile
with $n(r)\propto r^{-7/4}$, which is now commonly referred to as
the Bahcall-Wolf cusp. The extension of the cusp solution is given
by the radius of influence of the BH, $r_{i}$, which is defined as
\begin{equation}
r_{i}\equiv\frac{GM_{BH}}{2\,\sigma^{2}}\,\label{eq:r-influence}\end{equation}
where $G$ is the gravitational constant and $\sigma$ the velocity
dispersion of the core. As shown later by \citet{1976Natur.262..743S},
such a solution can be readily obtained using simple scaling arguments.
The key is to realize  that in the region delimited by the tidal radius,
$r_{t}$, within which stars get tidally disrupted, and $r_{i}$ within
which the stars are bound to the black hole, the net energy diffusion
timescale, $t_{U}$, is proportional to the local relaxation time,
$t_{r}$, which is the shortest timescale on which any physical quantity
can be transported. Furthermore,  the quasi-steady-state of the cusp
region is characterized by a dynamic equilibrium, with a constant
net energy flow into the core region  that should scale as $n(r)\, r^{3}E(r)/t_{U}$,
where $E(r)$ is the mean specific energy at radius $r$, so  $E(r)\sim GM_{BH}\, r^{-1}$
in the cusp region. Setting then $t_{U}\sim t_{r}\sim\sigma^{3/2}/\left[G^{2}m^{2}n(r)\right]\,$,
$\sigma\sim\sqrt{GM_{BH}\, r^{-1}}$, and simple  substitutions immediately
lead to $n(r)\propto r^{7/4}$. 

The formation of such a cusp has been confirmed subsequently by  numerical
studies. Most of them employed methods based on the Fokker-Planck
equation, solving it directly \citep{1976ApJ...209..214B,1977ApJ...211..244L,1978ApJ...226.1087C},
or indirectly, either using the statistical Monte Carlo approach \citep{1978ApJ...225..603S,2002A&A...394..345F}
or a fluid-dynamical approach based on velocity moments \citep{2004MNRAS.352..655A}.
Being derived from the Fokker-Planck equation, all these methods share
essentially the same set of underlying assumptions: (i) the cluster
potential has spherical symmetry; (ii) the cluster is in dynamical
equilibrium at all times; (iii) the evolution is driven by diffusive
2-body relaxation. Through direct $N$-body simulations that do not
rely on any \emph{a priori }assumptions, \citet{2004ApJ...613.1133B}
confirmed the cusp solution of \citet{1976ApJ...209..214B}, therefore
also providing important justification to the Fokker-Planck approach.

Based on the cusp solution, \citet{1976MNRAS.176..633F} calculated
the stellar disruption rate taking into account that stars inside
a critical radius, $r_{crit}$, are efficiently accreted by the black
hole as they diffuse quickly into low angular momentum orbits with
periastron distances, $r_{peri}$, smaller than $r_{t}$. As for a
given radial position, the velocity vectors that lead to orbits with
$r_{peri}<r_{t}$, form a cone, these orbits are also called loss-cone
orbits. Outside $r_{crit}$, stars are always able to leave the loss-cone
during one orbital period due to two-body relaxation while inside
$r_{crit}$ the orbital changes are smaller so that stars on loss-cone
orbits are more likely to reach the tidal radius and get disrupted
before they have a chance to get scattered out. Consequently, inside
$r_{crit}$ the loss-cone should be nearly empty while outside $r_{crit}$
it always remains full. In addition, \citet{1976MNRAS.176..633F}
argue that the disruption rate is mainly given by the cluster conditions
at $r_{crit}$. This is because, on the one hand, for $r>r_{crit}$
the fraction of stars populating the loss-cone decreases as the loss-cone
angle decreases with increasing radius, while for $r<r_{crit}$ the
total net flux of stars, and therefore the flux of stars that can
diffuse into the loss-cone, decreases rapidly \citep{1977ApJ...211..244L}.
Calculations by \citet{2004MNRAS.352..655A}  confirm the loss-cone
picture, showing that, for a cluster in dynamical equilibrium, the
disruption rate is strongly peaked at $r_{crit}$, and the fraction
of stars on loss-cone orbits is rapidly approaching zero for $r<r_{crit}$,
while for $r>r_{crit}$ it is always close to one. Similarly, \citet{2004ApJ...613.1133B}
find generally good agreement between the disruption rates in their
simulations and the disruptions rates based on the approximate expression
of \citet{1976MNRAS.176..633F} (their equation 22) and using the
cluster conditions at $r_{crit}$ from their $N$-body models.

While earlier work on the dynamics of clusters with IMBHs mainly focused
on the equilibrium state of the cusp surrounded by a static isothermal
core, \citet{1977ApJ...217..281S} considered the effect of an IMBH
on the global evolution of the cluster. Using a homological model
for the dynamical evolution, he calculated the core size in response
to evaporation of high-velocity stars and tidal disruption of stars
tightly bound to the IMBH. While stars that leave the cluster by evaporation
carry away very little energydriving core contraction \citep{1966ApJ...143..400S},
which would ultimately lead to core collapse in the absence of an
IMBH, the tidal disruptions close to the central IMBH that remove
stars with highly negative specific energies, provide an energy source
that causes the core to expand. \citet{1977ApJ...217..281S} shows
that for low initial IMBH masses and large initial core radii, stellar
evaporation first dominates and drives core contraction until, due
to the increasing core density, the tidal disruption rate becomes
large enough to reverse core collapse. The time of this reversal roughly
coincides with the time of core collapse for the cluster without IMBH.
Tidal disruptions then drive the re-expansion of the core, and the
core size increases asymptotically to infinity. This expansion is
a generic feature of a stellar system where energy is generated within
a very small central volume and the mass contained within this region
is very small compared to the cluster mass \citep{1965AnAp...28...62H,1977ApJ...217..281S}. 

The qualitative behavior of the core size evolution for lower mass
IMBHs was later confirmed by numerical studies \citep{1980ApJ...239..685M,1991ApJ...370...60M,2002A&A...394..345F,2004MNRAS.352..655A}.
\citet{2004MNRAS.352..655A} in particular showed that the core size
increases asymptotically as $\propto t^{2/3}$, which was also predicted
by \citet{1977ApJ...217..281S}. This expansion also causes the disruption
rate to decrease with time as approximately $\propto t^{-6/5}$ \citep{2004MNRAS.352..655A},
and will ultimately lead to the complete dissolution of the cluster
as the outer stars are removed by tidal forces \citep{1971Ap&SS..13..300W}.
For larger IMBH masses and small initial core sizes, tidal disruptions
will prevent any initial core contraction and the core expands from
the beginning \citep{1977ApJ...217..281S}. This case was calculated
by \citet{2004ApJ...613.1133B} using direct $N$-body simulations,
confirming that core expansion starts almost immediately and follows
a $t^{2/3}$ power-law. 

Most of the studies mentioned so far considered the evolution of cluster
containing a central massive black hole and comprised of stars of
equal mass. A stellar mass spectrum was first considered by \citet{1977ApJ...216..883B}
extending their previous work in \citet{1976ApJ...209..214B}. They
find that, due to mass-segregation, lower mass stars have shallower
density profiles than more massive ones. For old globular clusters
this means that the observable surface brightness cusp must be much
shallower as the more massive dark stellar remnants are concentrated
towards the center while the lower-mass main sequence stars that contribute
most of the light are much less centrally concentrated. It follows
that, although a cusp in the velocity and density profile provides
strong evidence for the presence of an IMBH in a cluster, such cusps
might not be easily detectable in real star clusters. Using direct
$N$-body simulations including an initial stellar mass spectrum and
stellar evolution, \citet{2004ApJ...613.1143B} find, indeed, flat
luminosity density profiles almost indistinguishable from a standard
King profile. Carrying out similar simulations but with $M_{BH}/M_{c}<1\%$,
\citet{2005ApJ...620..238B} find surface brightness cusps with power-law
slopes ranging from $\alpha=-0.1$ to $\alpha=-0.3$. Based on these
results they identified 9 candidate clusters from the sample of galactic
GCs of Noyola \& Gebhardt (2006) that might contain IMBHs.

Monte Carlo simulations of realistic clusters with central IMBH were
mainly done in the context of galactic nuclei. A recently developed
and well tested code is that of \citet{2002A&A...394..345F}. Similar
to our code, it is based on the method of \citet{1971Ap&SS..14..151H}
but is modified to evolve each star individually on a fixed fraction
of the local relaxation time, as opposed to the original shared time-step
scheme. In addition to the implementation of loss-cone physics, which
we will describe in detail in \S\ref{sec:Monte-Carlo-Method}, it
also incorporates stellar collisions interpolating between results
from detailed hydrodynamical simulations. Collisions between stars
is an important physical process in dense galactic nuclei. As already
pointed out by \citet{1976MNRAS.176..633F}, the radius $r_{coll}$
outside which stellar encounters responsible for relaxation can be
treated as elastic encounter can be larger than $r_{crit}$ for large
$M_{BH}$ and typical sizes and densities for galactic nuclei. Inside
$r_{coll}$ stars cannot deflect each other significantly without
colliding. As the relative velocity within $r_{coll}$ is larger then
the escape velocity from the stellar surface the collision can be
very disruptive and might under certain conditions provide a significant
source to fuel an active galactic nucleus \citep[see ][and references therein]{2002A&A...394..345F}.
However, for conditions typical inside globular clusters stellar collisions
are unlikely to play a significant role and are therefore not further
considered for the present study.

In addition to the formation of cuspy profiles, an IMBH influences
the surface brightness profile by producing rather large cluster cores
as measured by the core-to-half light radius, such that larger cores
are produced by more massive IMBHs \citep{2007PASJ...59L..11H}. The
large core sizes are simply a result of the energy flow from the central
cusp region to the core which causes the cluster to expand. Constructing
generalized King models \citep{1966AJ.....71...64K} including the
effect of an IMBH and a stellar mass spectrum, \citet{2007MNRAS.381..103M}
finds that the core size and the cusp slope are related such that
clusters with larger slopes, $s$, have lower concentrations, $c=\log(r_{c}/r_{t})$,
where $r_{c}$ is the core radius of the cluster. More specifically
they find that $s$ and $c$ are related by\begin{equation}
11.6s-4.85\lesssim\log\left(\frac{M_{BH}}{M_{c}}\right)\lesssim-1.14c-0.694\,,\label{eq:miocci}\end{equation}
where $M_{c}$ is the cluster mass. Based on this criterion and data
from \citet{2006AJ....132..447N} for $s$ as well as the Harris catalog
for $c$, they identified $7$ candidate clusters that might contain
IMBHs with mass $>100\,{\rm M}_{\odot}$, with 4 of them also identified
  by \citet{2005ApJ...620..238B}.

In contrast to the rather shallow surface brightness cusps, the stronger cusp
in the stellar velocity dispersion appears to be a much better diagnostic to
infer the presence of IMBHs in globular clusters. However, for globular
clusters this signature turns out to be difficult to detect as any velocity
dispersion measurement inside such a cusp has to rely on only a few bright
stars for expected IMBH masses $\lesssim1000\, M_{\odot}$ and typical globular
cluster masses \citep{2005ApJ...620..238B}.  These IMBH mass estimates are
based on extrapolating the well known $M_{BH}-\sigma$ relation
\citep{MagorrianRelation,GebhardtMsigma,FerrareseMerritt} between central black
hole mass and velocity dispersion in the central bulges of  galaxies down to
velocity dispersions typical for globular clusters ($\sim10\,{\rm km\,
s}^{-1}$). In that case $M_{BH}$ should be  $\sim10^{3}-10^{4}\,{\rm
M}_{\odot}$. Furthermore, simulations of collisional runaways  by
\citet{2004ApJ...604..632G} show that the mass of the Spitzer unstable
subcluster which provides the mass reservoir for forming the BH progenitor is
$\sim10^{-3}\,{\rm M}_{c}$, implying IMBH masses of at most a few $10^{3}\,{\rm
M}_{\odot}$ for typical cluster masses. However, it is important to point out
that this does not mean that $M_{BH}/M_{c}$ for an old globular cluster has to
be always significantly less than $1\%$, as a cluster can later loose a
substantial amount of mass due to tidal stripping in the Galactic field.

\section{Method and Initial Conditions\label{sec:Monte-Carlo-Method}}

\subsection{Monte Carlo Method with IMBH}

Our MC code shares some important properties with direct $N$-body
methods, which is why it is also regarded as a randomized $N$-body
scheme \citep[see, e.g.,][]{2001A&A...375..711F}. Just as in direct
$N$-body codes, it relies on a star-by-star description of the GC,
which makes it particularly straightforward to include additional
physical processes such as stellar evolution. Contrary to direct $N$-body
methods, however, the stellar orbits are resolved on a relaxation
time scale $t_{r}$, which is much larger than the crossing time $t_{cr}$,
the time scale on which direct $N$-body methods resolve those orbits.
The specific implementation we use for our study is the MC code initially
developed by \citet{2000ApJ...540..969J} and further enhanced and
improved by \citet{2003ApJ...593..772F} and \citet{2007ApJ...658.1047F}.
The code is based on Hénon's algorithm for solving the Fokker-Planck
equation. It incorporates treatments of mass spectra, stellar evolution,
primordial binaries, and the influence of a galactic tidal field.

\begin{figure}
\includegraphics{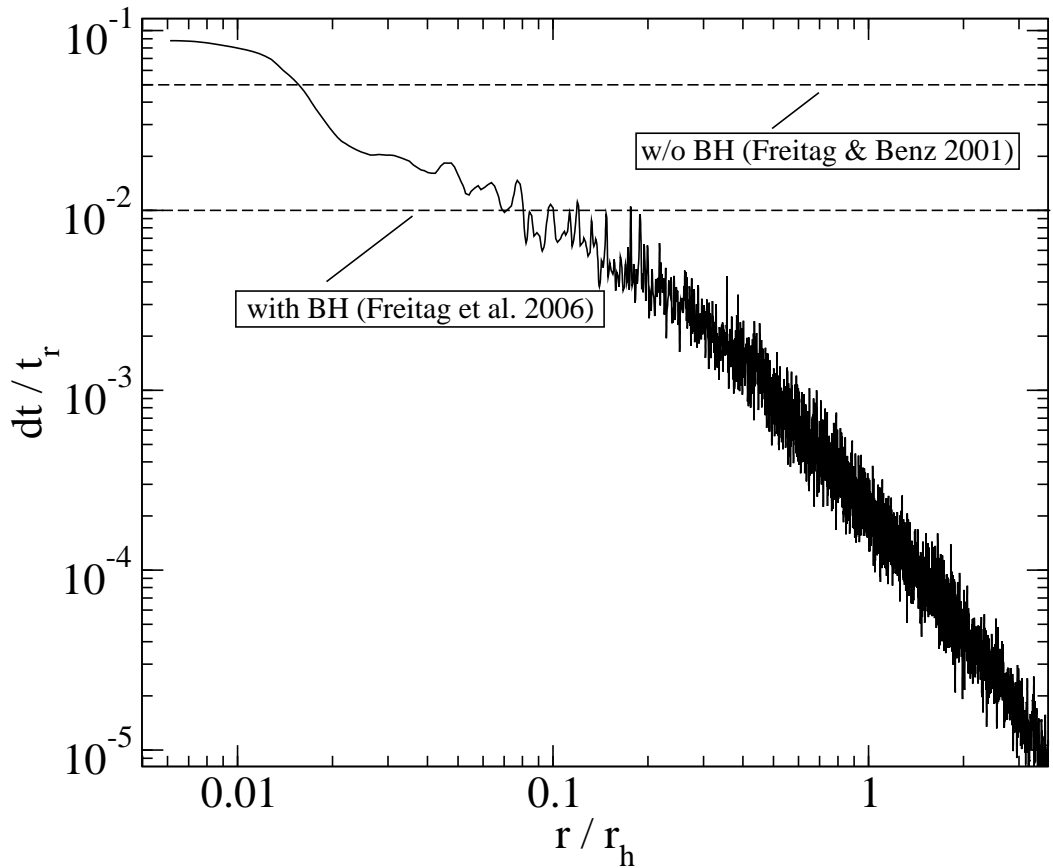}

\caption{\label{fig:timestep}Snapshot of the ratio of global MC timestep to
local relaxation time. While in the individual timestep scheme of
\citet{2002A&A...394..345F} $dt$ is a constant fraction of $t_{r}(r)$
(dashed lines), in our shared timestep scheme $dt/t_{r}(r)$ (solid
line) is decreasing with increasing $r$ as $t_{r}$ increases. As
$dt$ must be chosen large enough to avoid artificial relaxation in
the shared timestep scheme, $f$ is larger for $r<r_{h}$ than in
\citet{2006ApJ...649...91F}, resulting in a slower expansion of the
cluster.}

\end{figure}

The effect of an IMBH on the stellar distribution is implemented in
a manner similar to that of \citet{2002A&A...394..345F}. In this
method the IMBH is treated as a fixed, central point mass while stars
are tidally disrupted and accreted onto the IMBH whenever their periastron
distances lie within the tidal radius, $r_{t}$, of the IMBH, which
is given by \begin{equation}
r_{t}=\left(2\frac{M_{BH}}{M_{*}}\right)^{1/3}R_{*}\,,\label{eq:rtidal}\end{equation}
 where $R_{*}$ and $M_{*}$ are the stellar radius and mass respectively.
Stars are removed from the system and their masses are added to the
BH as soon as their velocity vector, v, enter the loss-cone, $\theta_{LC}$,
approximately given by\[
\theta_{LC}^{2}\simeq2\frac{GM_{BH}r_{t}}{v^{2}r^{2}}\,.\]
However, as the star's removal happens on an orbital timescale one
would need to use timesteps as short as the orbital period of the
star in order to treat the loss-cone effects in the most accurate
fashion. This would, however, slow down the whole calculation considerably.
Instead, during one MC timestep a star's orbital evolution is followed
by simulating the random-walk of its velocity vector, which approximates
the effect of relaxation on the much shorter orbital timescale. The
random-walk procedure goes as follows:
\begin{enumerate}
\item After a gravitational encounter between two stars is calculated in
the standard Monte Carlo fashion, resulting in a deflection angle
of $\delta\theta_{step}$, the orbital period, $P_{orb}$, is calculated
using Gauss-Chebychev quadrature, and a {}``representative'' diffusion
angle during a single orbit, $\delta\theta_{orb}$, is estimated as
\[
\delta\theta_{orb}=\left(\frac{\delta t}{P_{orb}}\right)^{-\frac{1}{2}}\delta\theta_{step}\]
where $\delta t$ is the time step. 
\item The star's velocity vector with respect to the encounter reference
frame is calculated and a variable $L_{2}$, which represents the
remaining quadratic deflection, is set to $\delta\theta_{step}^{2}$.
\item The star is tested for entry into the loss-cone, and , if this is
case, is removed and its mass added to $M_{BH}$, whereupon the random-walk
is terminated. Otherwise, we proceed with the next step.
\item If $L_{2}\leq0$, the random-walk is terminated. The star's position
and velocity are reset to its values before the random-walk procedure.
\item A new random-walk step is carried out with amplitude $\Delta=\max\left(\delta\theta_{orb},\min\left(0.1\pi,\Delta_{safe},\sqrt{L_{2}}\right)\right)$,
and a random direction on the velocity sphere, where $\Delta_{safe}$
is set to roughly half the angular distance to the loss-cone. This
way, $\Delta$ becomes progressively smaller down to $\delta\theta_{orb}$
when approaching the loss-cone in order to keep the risk of missing
a disruption minimal. According to the new step size, the direction
of the star's velocity is changed, and $L_{2}$ is updated: $L_{2}:=L_{2}-\Delta$.
The random-walk continues at step 3.
\end{enumerate}
Although many of our results turn out to agree very well with previously
published data, there are discrepancies when comparing the disruption
rates. One possible reason for these differences might be related
to fact that our code uses a shared timestep scheme. In an individual
timestep scheme, as in \citet{2001A&A...375..711F}, the timestep
is some constant fraction of the local relaxation time, i.e., $dt_{i}=f~t_{r}(r_{i})$,
where $f$ is a constant,and the subscript $i$ refers to the individual
star. In a shared time step scheme the smallest of these $dt_{i}$
is chosen for all stars. This results in much shorter time steps for
stars further out in the cluster compared to an individual timestep
scheme. As has been noted by \citet{2006ApJ...649...91F}, the timestep
size must be chosen small enough in order  to achieve good agreement
with $N$-body simulations,  with $f\lesssim0.01$. While choosing
such a small time-step was still feasible in the code of \citet{2002A&A...394..345F},
to enforce such a criterion for all stars in our code would lead to
a dramatic slow-down and notable spurious relaxation as the timesteps
for the stars in the outer cluster regions relative to the local relaxation
time become extremely small \citep[see][]{2004ApJ...604..632G}. In
order to reduce the effect of spurious relaxation we are forced to
choose a larger $dt$ resulting in a larger $f$ for the inner regions,
up to $f\approx0.1$. Fig. \ref{fig:timestep}%
 shows an example of $dt/t_{r}(r)$ as a function of radius. One can
see that in this case only for $r\gtrsim0.1\, r_{h}$ we have $dt/t_{r}(r)\lesssim0.01\,$,
while for the inner region it rises quickly to $0.1$. As a result,
any expansion of the inner cluster region, which limits the growth
of the IMBH in our comparison calculations, will be slower and the
disruption rates, therefore, larger, due to the higher core densities.
Such an expansion occurs either in response to the growing IMBH mass
or as an initial expansion of the cluster. Previously, we applied
a procedure that tries to compensate for the larger time steps in
the inner region \citep{2008IAUS..246..351U} by evolving the inner
stars individually on smaller timesteps while keeping the cluster
potential constant during one shared time step, a procedure that bears
some resemblance to the method of \citet{1980ApJ...239..685M}. However,
it turned out that in order to achieve good agreement with direct
$N$-body simulations one has to choose the timestep sizes for each
cluster configuration separately. This is not only undesirable but
might also imply that there are additional processes at work that
significantly influence the disruption rate and are not included in
the Monte Carlo scheme. In \S\ref{sub:Comparison-with-{Baumgardt}}
we discuss several possibilities, among them the wandering of the
IMBH. As there is no obvious way to compensate for such a processes
in a uniform and consistent way through adjustments in the two-body
relaxation time scale, we do not use the sub-timestep scheme for the
present paper.

\subsection{Initial Conditions}

Table \ref{tab:mc-bme} %
\begin{sidewaystable}
\begin{tabular}{|c|c|c|c|c|>{\centering}p{1.5in}|c|>{\centering}p{1.5in}|}
\hline 
Name & $N$ & $W_{0}$ & $r_{t}\,[R_{vir}]$ & $M_{BH,i}\,[M_{\odot}]$  & $M_{BH,f}\,[M_{\odot}]$ & $T_{END}\,[t_{cr}]$ & $N_{Disr}$\tabularnewline
\hline
\hline 
BME-1 & $80,000$ & 10 & $1\times10^{-7}$ & $266$ & $1429\pm35\,(827)$ & $3000$ & $1163\pm35\,(561)$\tabularnewline
\hline 
BME-2 & $80,000$ & 10 & $1\times10^{-7}$ & $800$ & $1690\pm20\,(1388)$ & $2000$ & $890\pm20\,(588)$\tabularnewline
\hline 
BME-3 & $80,000$ & 10 & $1\times10^{-7}$ & $2660$ & $3434\pm20\,(3285)$ & $2000$ & $774\pm20\,(625)$\tabularnewline
\hline 
BME-16 & $178,800$ & 10 & $1\times10^{-7}$ & $461$ & $2171\pm30\,(1368)$ & $2000$ & $1710\pm30\,(907)$\tabularnewline
\hline 
ASFB-50 & $100,000$ & Plummer & $2.26\times10^{-8}$ & $50$ & $10500\pm100\,$ $(7450,13050)$ & $5.4\times10^{6}$ & $10,450\pm100$ $(7.4\times10^{3},\,1.3\times10^{4})$\tabularnewline
\hline 
ASFB-500 & $100,000$ & Plummer & $2.26\times10^{-8}$ & $500$ & $11500\pm100$ $(8900,14500)$ & $5.4\times10^{6}$ & $11,000\pm100$ $(8.4\times10^{3},\,1.4\times10^{4})$\tabularnewline
\hline
\end{tabular}

\caption{\label{tab:mc-bme}Details of the performed Monte Carlo runs for single-mass
clusters with initial conditions as in \citet{2004ApJ...613.1133B}
(BME) and \citet{2004MNRAS.352..655A} and \citet{2002A&A...394..345F}
(ASFB). Values in parentheses are results from the corresponding literature.
Where two values are given the first one refers to \citet{2002A&A...394..345F}
and the second to \citet{2004MNRAS.352..655A}. }

\end{sidewaystable}
\begin{sidewaystable}
\begin{tabular}{|c|c|c|>{\centering}p{0.8in}|c|>{\centering}p{0.75in}|c|>{\centering}p{0.78in}|>{\centering}p{0.75in}|>{\centering}p{0.8in}|>{\centering}p{0.7in}|}
\hline 
Name & $N$ & $W_{0}$ & $M_{BH,i}\,[M_{\odot}]$  & $r_{h,i}[{\rm pc}]$ & $M_{BH,f}\,[{\rm M}_{\odot}]$ & $N_{Disr}$ & $M_{c,f}\,[{\rm M}_{\odot}]$ & $T_{END}\,[{\rm Gyr}]$ & $r_{h,f}\,[{\rm pc}]$ & $\log t_{rh}\,[{\rm yr}]$\tabularnewline
\hline
\hline 
BMH-1 & $131,072$ & 7 & 125 & $4.91$ & $133\pm3$ (137) & $4\pm2$  & $45,671\pm6$ ($45,534$) & 12 & $12.02\pm0.05$ ($12.31$) & $9.86$ ($9.82$)\tabularnewline
\hline 
BMH-2 & $131,072$ & 7 & 250 & $4.91$ & $260\pm4$ (280) & $6\pm2$  & $45,677\pm14$ ($45,311$) & 12 & $11.98\pm0.04$ $(12.60)$ & $9.86$ ($9.84$)\tabularnewline
\hline 
BMH-3 & $131,072$ & 7 & 500 & $4.91$ & $513\pm5$ (531) & $9\pm3$  & $45,400\pm30$ ($44,741$) & 12 & $12.36\pm0.04$ ($13.70$) & $ $$9.88$ ($9.89$)\tabularnewline
\hline
\end{tabular}

\caption{\label{tab:mc-bmh}Details of the performed Monte Carlo runs for multi-mass
clusters with initial conditions as in \citet{2005ApJ...620..238B}
(BMH). The star masses were chosen according to a Kroupa mass function
ranging from $0.1-30\,{\rm M}_{\odot}$. Stellar evolution was modeled
using the SSE code of \citet{2002MNRAS.329..897H}. }

\end{sidewaystable}
 summarizes the initial conditions and main results for all our single-mass
runs and Table \ref{tab:mc-bmh} for our multi-mass runs that include
stellar evolution  (implemented by \citealt{2009ApJ...695L..20F}
using the SSE code of \citealt{2002MNRAS.329..897H}). 

The single-mass clusters consist of $N$ stars all of mass $1\,{\rm M}_{\odot}$
with positions and velocities chosen according to a Plummer model
or a King model with dimensionless potential $W_{0}=10$. Radii and
times are given in terms of the virial radius, $R_{vir}$, and the
crossing time, $t_{cr}$, respectively, which are defined by \begin{equation}
R_{vir}=\frac{GM_{c}^{2}}{-4\, E_{0}}\,,\end{equation}
and \begin{equation}
t_{cr}=\frac{GM_{c}^{5/2}}{\left(-4E_{0}\right)^{3/2}}\,.\end{equation}
where $E_{0}$ is the total gravitational energy of the cluster. The
cluster was evolved up to a time $T_{END}$, with an initial IMBH
mass $M_{BH,i}$ and $r_{t}$ a constant for all stars. Also shown
are the resulting final IMBH mass, $M_{BH,f}$ and the number of tidally
disrupted stars, $N_{disr}$. 

The initial conditions for multi-mass clusters are chosen as in BMH,
with stellar masses drawn from a Kroupa mass function \citep{2001MNRAS.322..231K}
in the range of $0.1-30\,{\rm M}_{\odot}$, stars evolved at a metallicity
$Z=0.001$, and initial positions and velocities chosen according
to a King model with $W_{0}=7$. The tidal disruption radius for each
star is given by equation \ref{eq:rtidal} with stellar mass and radius
provided by the stellar evolution code. In addition to $M_{BH,i}$,
$M_{BH,f}$ and $N_{disr}$ the half-mass radius, $r_{h}$, cluster
mass, $M_{c}$, and the final relaxation time at the half-mass radius,
$t_{rh}$, are shown, where the added subscripts $i$ and $f$ indicate
initial and final values, respectively.

For each set of initial conditions $9$ Monte Carlo runs were performed
with varying random seed, and the values given in the table are the
averages and standard deviations of these runs.

\section{Idealized Models\label{sec:Idealized-Models}}

\subsection{Imprints of IMBHs}

\label{ImprintsIMBH} 
\begin{figure}
\includegraphics[height=0.8\textheight]{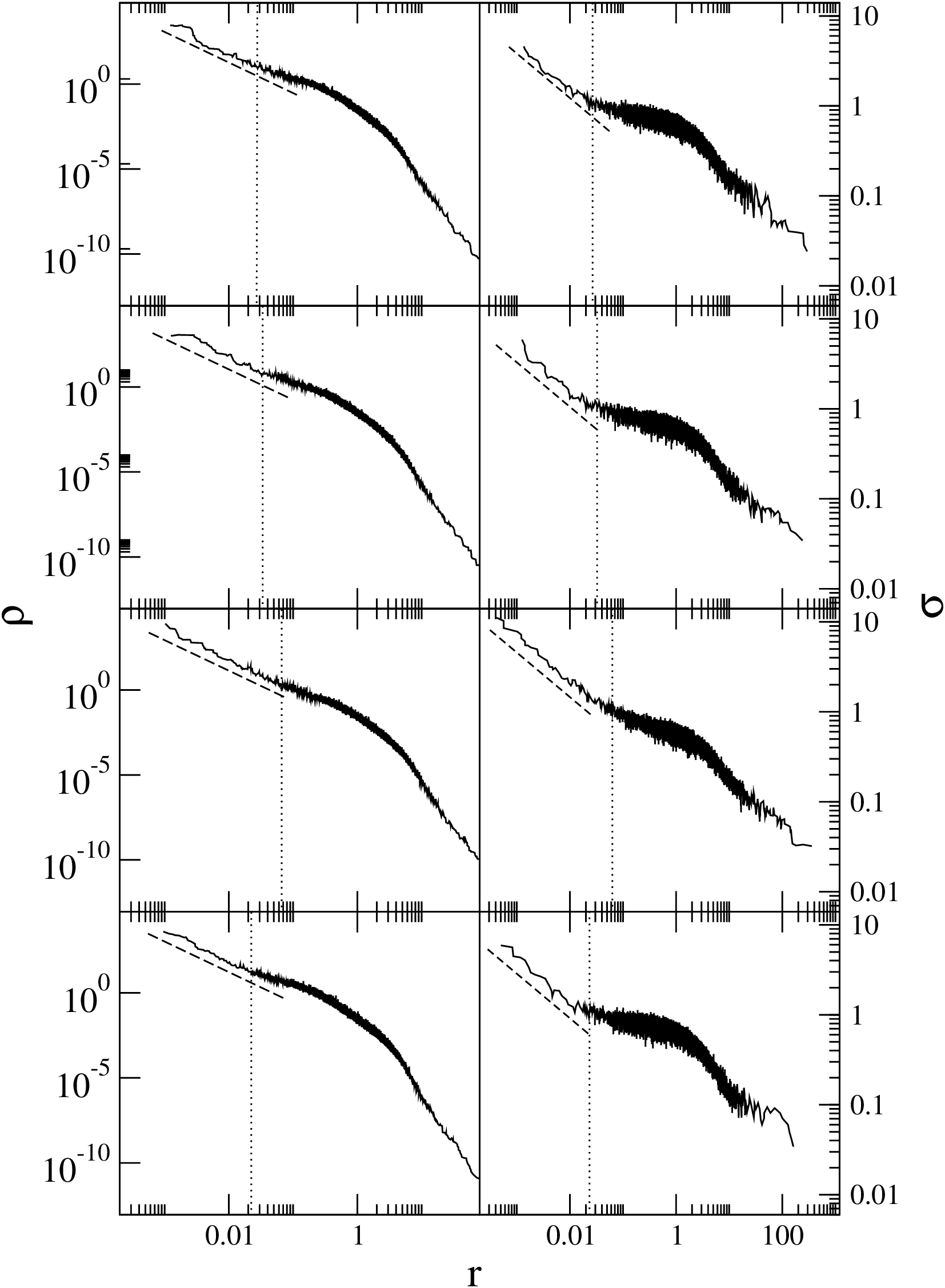}

\caption{\label{Imprints}Density, $\rho(r)$, (left panels) and velocity dispersion
profiles, $\sigma(r)$, (right panels) for runs BME-1 to BME-3 and
BME-16 from top to bottom, respectively. $\rho$ is given in units
of $M_{c}/R_{vir}^{3}$ and $\sigma$ in units of $R_{vir}/t_{cr}$.
The dotted line indicates the radius of influence of the IMBH and
the dashed lines represent the theoretically expected power-laws scalings
$\rho\sim r^{-7/4}$ or $\sigma\sim r^{-1/2}$.}

\end{figure}

In Fig. \ref{Imprints} %
density and velocity profiles from our simulations of single-mass
clusters are shown, together with the expected $r_{i}$.  calculated
from equation (\ref{eq:r-influence}) and using $\sigma\approx0.55$
in $N$-body units, appropriate for a $W_{0}=10$ King model. As can
be clearly seen, the density profile of the inner region of the evolved
clusters follow closely the expected $n(r)\propto r^{-7/4}$ power-law
and the extent of the cusp matches that of the region where the velocity
dispersion is Keplerian, which in turn matches the expected $r_{i}$.
However, contrary to what is seen in direct $N$-body simulations,
the cusp extends down to much smaller radii, especially for black
hole masses below $1\%$ of the cluster mass. This is mainly because,
in our simulations, the central IMBH has a fixed position, while in
direct $N$-body simulations it is allowed to move freely. As a consequence,
the density profile flattens inside its wandering radius compared
to a pure cusp profile, resulting in fewer stars in the central region.
As will be shown later, this might have an influence on the rate at
which stars are tidally disrupted.

\subsection{Disruption Rates}

\subsubsection{Comparison with \citet{2004MNRAS.352..655A} and \citet{2002A&A...394..345F}\label{sub:Comparison-with-{Amaro-Seoane}}}

\begin{figure}
\includegraphics[clip,width=0.8\textwidth]{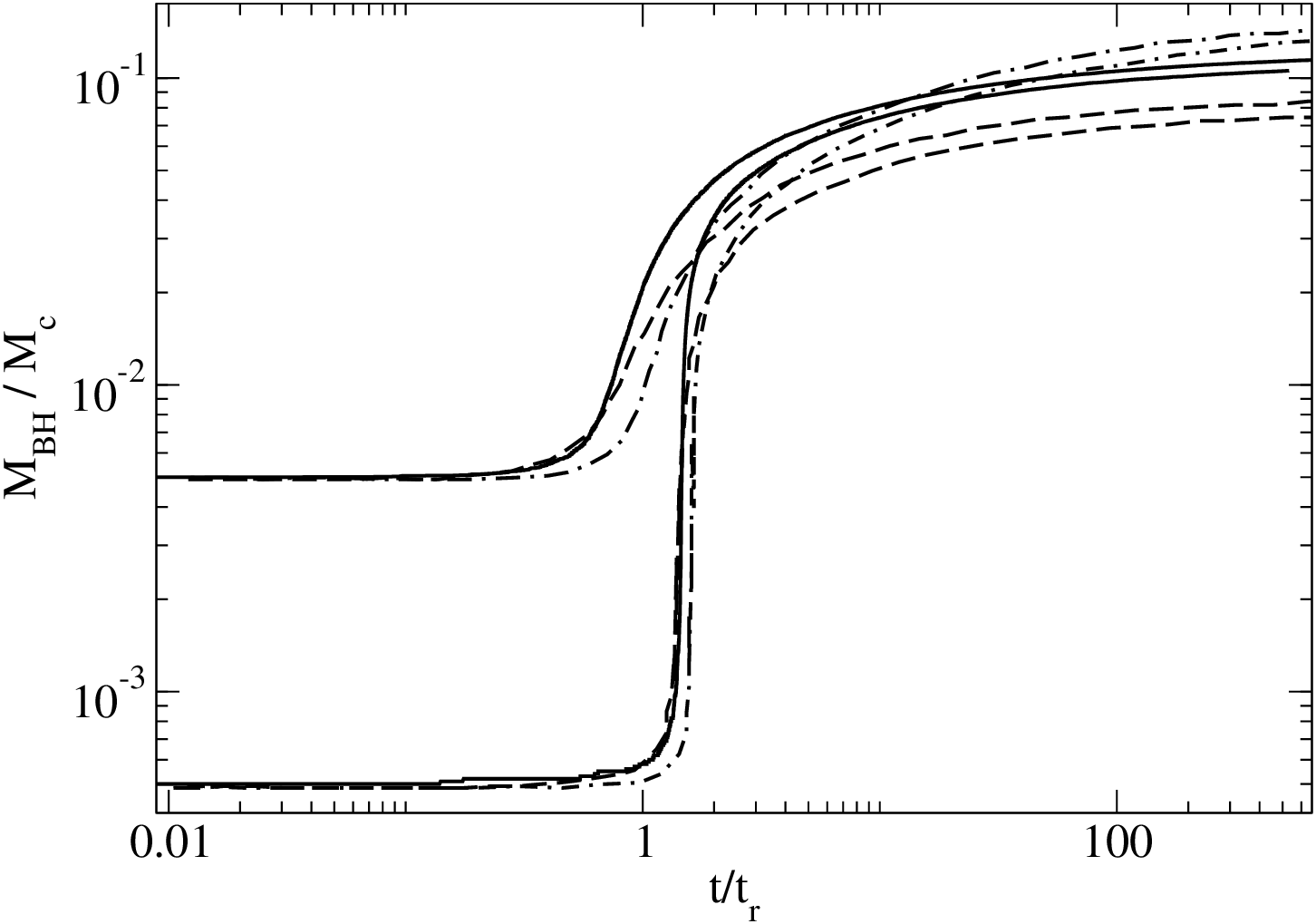}

\caption{\label{fig:AS01}BH mass as a function of time for two different initial
BH masses, $50\,{\rm M}_{\odot}$ (lower set) and $500\,{\rm M}_{\odot}$
(upper set). Shown are results from the gas code of \citet{2004MNRAS.352..655A}
(dash-dotted lines) and the Monte Carlo codes of \citet{2002A&A...394..345F}
(dashed lines) and ours (solid lines). }

\end{figure}

\begin{figure}
  \includegraphics[clip,width=0.8\textwidth]{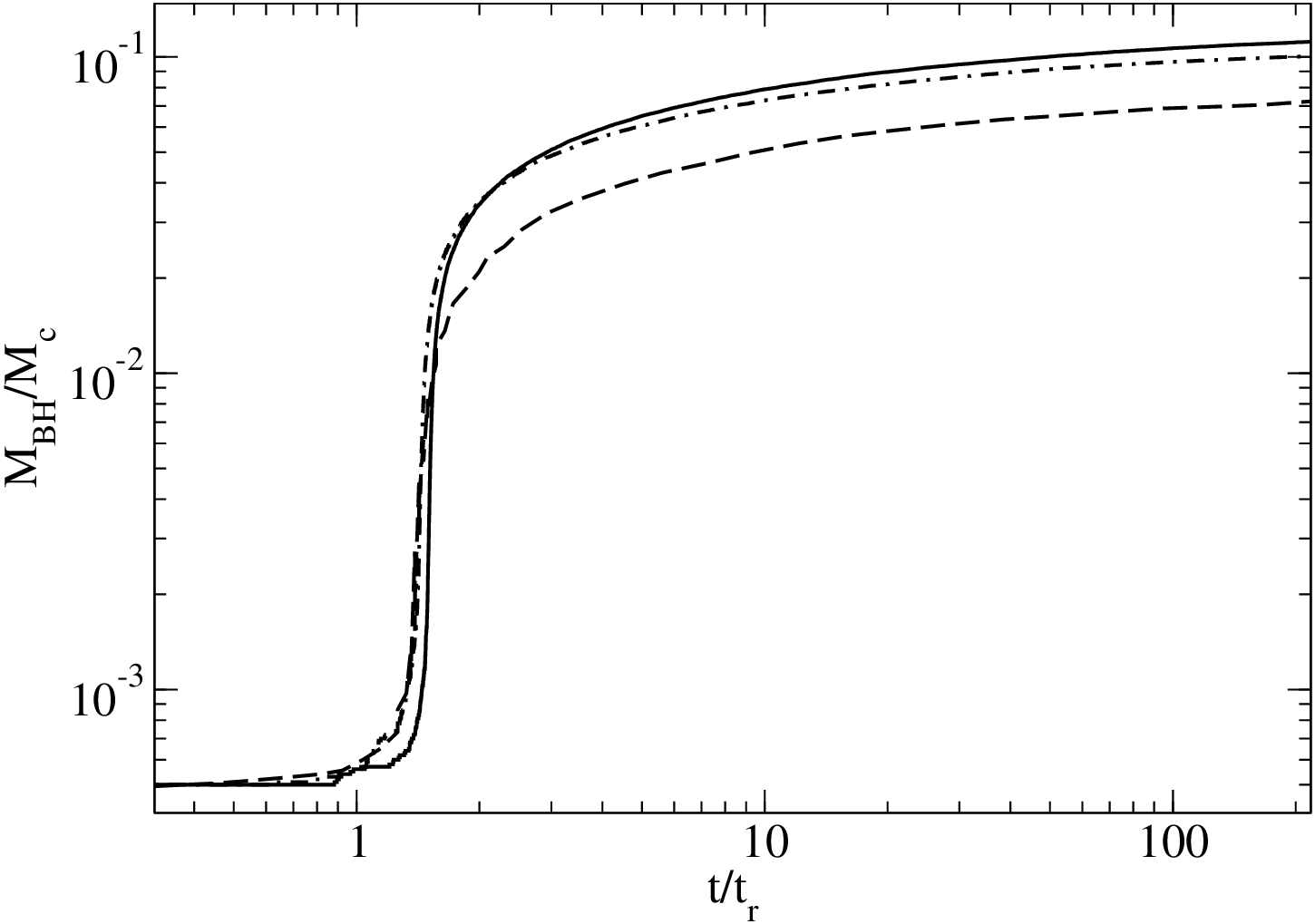}

\caption{\label{fig:FB02-dt}BH mass as a function of time for two values of 
the timestep parameter, $\theta_{max}=1$ (solid line) and $\theta_{max}=0.6$
(dash-dotted line). Also shown are results from the Monte Carlo code 
of \citet{2002A&A...394..345F} (dashed line). All calculations started with 
an IMBH mass of $50\,{\rm M}_{\odot}$. Larger timesteps cause a delay in the 
onset of the rapid growth phase, due to the slower core contraction, and slow 
down the core expansion of the cluster in response to the mass growth of the 
IMBH, leading to larger accretion rates.}

\end{figure}

Fig. \ref{fig:AS01}%
 compares the growth of the IMBH in our  simulations and those of
\citet{2004MNRAS.352..655A}, using a gas code, and \citet{2002A&A...394..345F},
using an individual time step Monte Carlo code. Here the evolution
of a cluster with $10^{5}$ stars all of $1\, M_{\odot}$ and a fixed
massive BH in the center with a mass of $50\,{\rm M}_{\odot}$ or
$500\, M_{\odot}$ was calculated. The stars were initially distributed
according to a Plummer density profile with $r_{h}=0.707\,{\rm pc}$
. As can be seen, similar to the results of the Monte Carlo code of
\citet{2002A&A...394..345F}, our results match qualitatively the
results of the gaseous method of \citet{2004MNRAS.352..655A} inasmuch
as there is a steep rise of $M_{BH}$ at the time when the cluster
densities are largest, which coincides with the time the cluster would
formally go into core collapse if there were no IMBH. After that time
the BH growth levels off due to the expansion of the core, described
in \S\ref{sec:Previous-Work-of}, leading to the convergence of the
IMBH mass to an asymptotic value. Quantitatively, however, there are
differences in  the onset of the rapid growth phase as well as in
the value of the final IMBH mass. The IMBH masses in \citet{2004MNRAS.352..655A}
are generally larger at late times, while the rapid growth phase is
somewhat delayed. Our results follow more closely, and unsurprisingly,
the ones by \citet{2002A&A...394..345F} up until shortly after the
onset of the rapid growth phase, while they converge to larger $M_{BH}$
at late times. Part of the reason for this discrepancy with \citet{2002A&A...394..345F}
must be related to the larger timestep size compared to the local
relaxation time in the inner region of the cluster and, thus, their slower
expansion (discussed in \S \ref{sec:Monte-Carlo-Method}). This is further
demonstrated in Fig. \ref{fig:FB02-dt}, where we plot the IMBH growth for two
different values of the timestep parameter $\theta_{max}$, which is the maximum
deflection angle for all stars \citep[see ][their eq. 7]{2001A&A...375..711F}.
Here we clearly see that increasing the timestep size increases the IMBH mass,
while the onset of the rapid IMBH mass growth is delayed. Both effects can be
ascribed to the system becoming more and more under-relaxed for larger
timesteps. First, the core contraction phase becomes longer, causing the delay
of the onset of the rapid growth phase, and, second, the cluster expands
slower, increasing the period of high core densities, and, thus, accretion rates, 
as discussed in \S\ref{sec:Monte-Carlo-Method}.
 
Despite these differences, the asymptotic IMBH masses differ by less than a
factor of two and are, therefore, in reasonable agreement with each other,
given the very long integration time.

\subsubsection{Comparison with BME\label{sub:Comparison-with-{Baumgardt}}}

We now compare the growth rate of IMBHs in our simulations with the
direct $N$-body simulations of BME. For this comparison we restrict
ourselves to runs with a larger number of particles to ensure that
the central cusp around the IMBH is sufficiently populated with stars
for the Monte Carlo method to be applicable (see Table \ref{tab:mc-bme}).

\begin{figure}
\begin{tabular}{cc}
\includegraphics[clip,width=0.45\columnwidth]{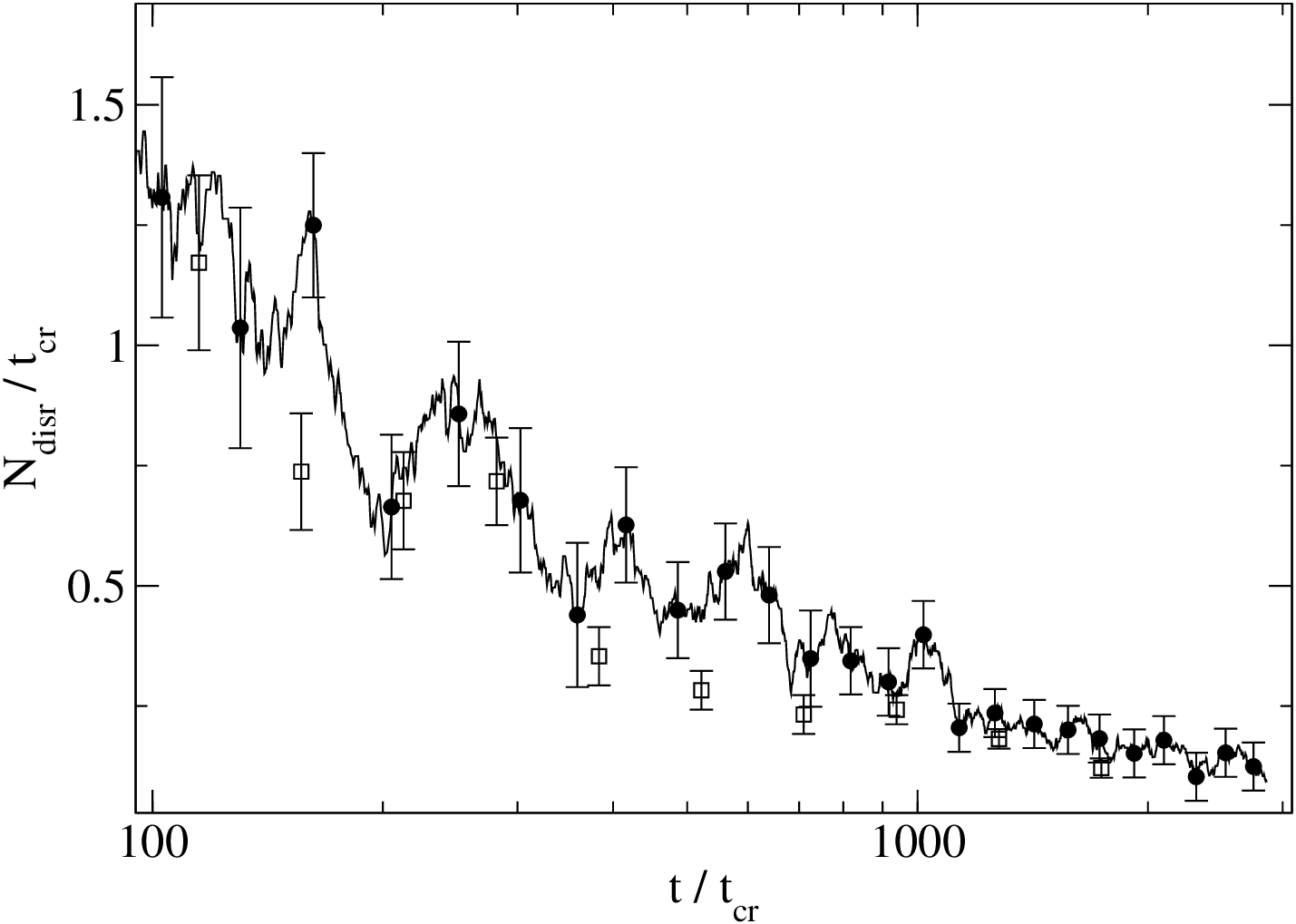} & \includegraphics[width=0.45\columnwidth]{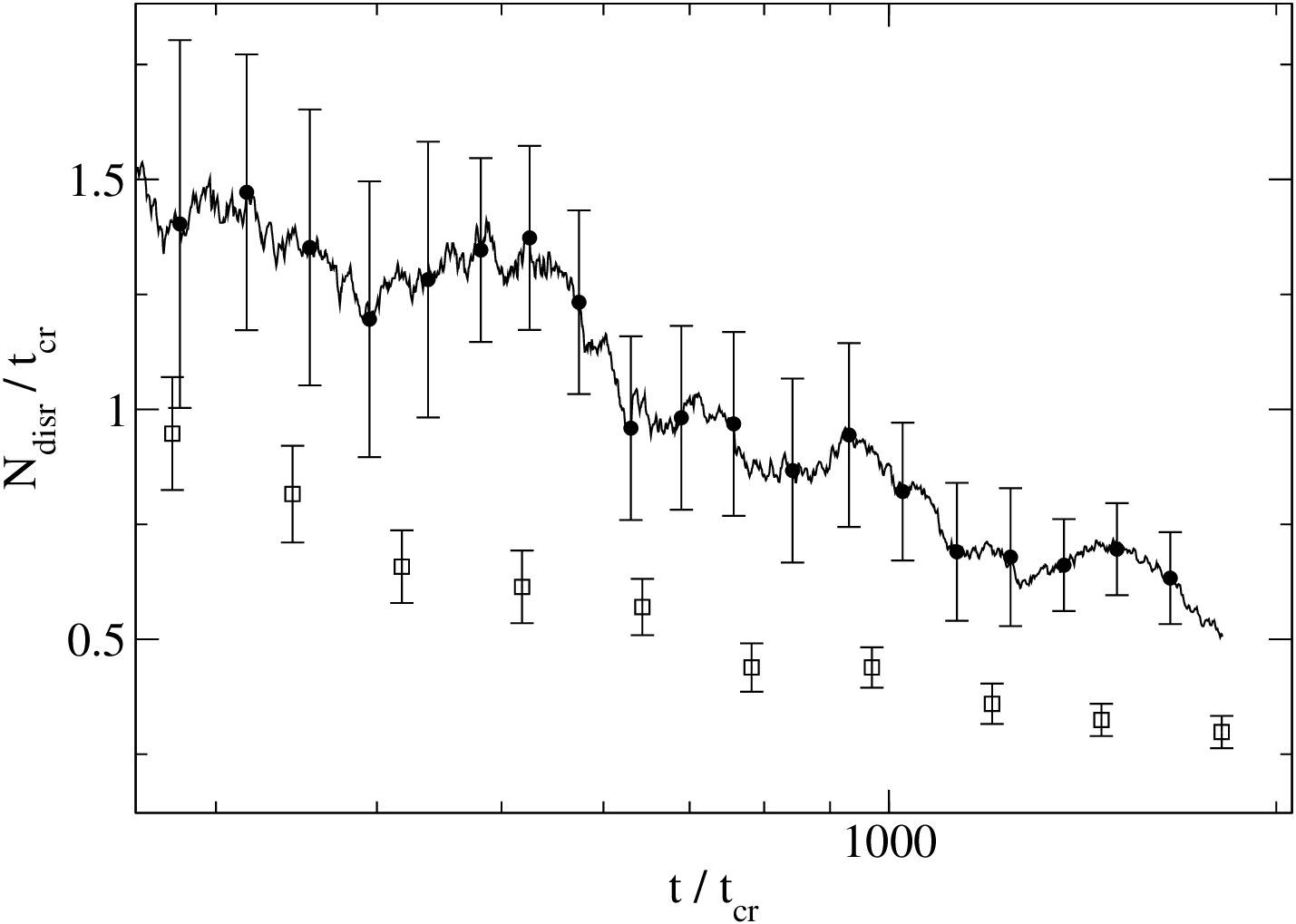}\tabularnewline
\end{tabular}

\caption{\label{fig:disrupt-no-substeps}Comparison of the disruption rates
per crossing time for the direct $N$-body results of \citet{2004ApJ...613.1133B}
(open symbols) and for our MC simulations (filled symbols). Shown
are results of the runs BME-2 (left) and BME-16 (right). The disruption
rates of the MC runs are the average rates from $9$ runs with different
random seeds, and the error bars are the standard deviation of this
average. The disruption rates and error bars of the $N$-body results
are time averages and their standard deviations respectively.}

\end{figure}

Fig. \ref{fig:disrupt-no-substeps}%
 shows the disruption rate as a function of time for runs BME-2 and
BME-16 for our Monte Carlo and the direct $N$-body code. The qualitative
behavior, i.e. the decrease of the disruption rate due to the expansion
of the cluster is well reproduced. However, our rates are systematically
larger, always leading to IMBH masses that are larger than in direct
$N$-body simulations (see Table \ref{tab:mc-bme}). We find the largest
discrepancies for the lowest $M_{BH}/M_{c}$ ratios, where the total
number of disruptions differ by a factor of $2$, whereas the disruption
rates differ by approximately a factor of $1.5$. The difference in
the disruption rates becomes quickly smaller for larger IMBH masses,
so that for $M_{BH}/M_{c}=1\%$ (Fig. \ref{fig:disrupt-no-substeps}
(left panel) there is agreement within the error bars. However, our
rates are still systematically larger for that case , leading to approximately
$50\%$ more   disruptions at the end of the simualtion    compared
to $N$-body simulations. This difference is again smaller for $M_{BH}/M_{c}\approx3\%$
going down to only $ $$25\%$. 

There are several possible reasons that may explain larger disruption
rates in our simulations compared to direct $N$-body simulations.
First, one has to note that our results in Fig. \ref{fig:disrupt-no-substeps}
are the averages over $9$ runs with different initial seeds to generate
the cluster, while the disruption rates of \citet{2004ApJ...613.1133B}
come from only one realization of a cluster and are time averages.
As the run-to-run variations seem to be larger than the time variations
in our MC runs, it might be possible that this also applies to the
$N$-body results, in which case the differences in disruption rates
could be statistically less significant or even disappear for larger
BH masses.

Furthermore, both calculations start from non-equilibrium cluster
configurations, consisting of a central massive IMBH and a flat constant
density core \citep{2004ApJ...613.1133B}. As the Monte Carlo method
assumes that the cluster is always in dynamic equilibrium, it cannot
adequately model the initial phase until such an equilibrium is reached.
It might, therefore, be possible that the difference in the number
of tidal disruptions is at least in part due to differences in modeling
the initial, violent relaxation process. On the other hand, the number
of disruptions during the first $100\, t_{{\rm cr}}$ account for
at most $5\%$ of the total in all runs, indicating that these differences
have a rather minor influence on our results.

Another effect is the wandering of the IMBH due to close passages
of stars that are not bound to the black hole. Although, one would
intuitively think that a wandering IMBH would increase the cross section
for stars to enter the tidal radius as it covers a larger volume and
thus provides a larger cross section \citep{2002ApJ...572..371C},
the wandering also tends to flatten the density profile \citep{2004ApJ...613.1133B,2004MNRAS.352..655A}.
As the disruption rate is proportional to the density near the IMBH,
either at $r_{crit}$ if the loss-cone is empty, or $r_{t}$ if the
 loss-cone is full, such a flattening of the density profile could,
therefore, cause a lower disruption rate. Similarly, \citet{1999MNRAS.309..447M}
discuss the influence of BH wandering on the disruption rate, arguing
that, depending on cluster parameters and BH mass, the disruption
rate can be increased or decreased relative to the disruption rate
of a fixed central BH.

In order to estimate this effect more quantitatively we assume for
simplicity that, in the case of a wandering IMBH, the IMBH remains
inside the cusp and the stellar density, $n(r)$, outside of the wandering
radius, $r_{w}$, is given by $n_{cusp}\sim r^{-7/4}$, while inside
it has a constant value of $n_{cusp}(r_{w})$. We furthermore make
the assumption that, in this case, we are in the  full-loss-cone regime,
i.e., the loss-cone is constantly replenished with stars on a dynamical
timescale, contrary to the empty-loss-cone regime, where this happens
on a relaxation time. The assumption of a full loss-cone for a wandering
IMBH is justified as long as its motion is fast enough so that the
influx of stars is always sufficiently high to completely replenish
the loss-cone. The situation is of course different for a fixed BH
where the loss-cone is not replenished but rather depleted on such
a short timescale. In effect, we are comparing the disruption rate
of fixed BH in the empty loss-cone regime with that of a wandering
BH in the full loss-cone regime.

The disruption rate for a wandering IMBH in the full loss-cone regime
is simply given by the product of the disruption cross section of
the IMBH, $\Sigma_{BH}$, the stellar number density, $n_{*}$, and
the relative velocity dispersion between IMBH and stars, $v_{rel}$,
evaluated at the wandering radius, $r_{w}$, as \citep[see][their eq. 8-123]{2008gady.book.....B}
\begin{equation}
\dot{n}_{w}=n_{*}(r_{w})\,\Sigma_{BH}(r_{t},r_{w})\, v_{rel}(r_{w})\end{equation}
where $\Sigma_{BH}$ is given by\begin{equation}
\Sigma_{BH}(r_{t},r_{w})=4\sqrt{\pi}r_{t}^{2}\left(1+\frac{GM_{BH}}{r_{t}\,2\, v_{rel}^{2}(r_{w})}\right)\,,\label{eq:cross-section-BH}\end{equation}
and $v_{rel}=\sqrt{(v_{*}^{2}+v_{BH}^{2})/2}$ with $v_{*}$ and $v_{BH}$
denoting the stellar and the IMBH velocity dispersion, respectively.
Using $v_{BH}\ll v_{*}$, $v_{*}(r)=\sqrt{\nicefrac{GM_{BH}}{r}}$
in the cusp region, $r_{w}\gg r_{t}$,we obtain \begin{equation}
\dot{n}_{w}=4\sqrt{\pi}\, n_{c}\, v_{c}\, r_{i}\left(\frac{r_{i}}{r_{w}}\right)^{5/4}r_{t}\label{eq:ndot-wandering}\end{equation}
where the subscript $c$ refers to the core, $n_{c}=n(r_{i})$, and
$v_{c}=v(r_{i})$. 

For a fixed IMBH the disruption rate is given by the influx of stars
into the region inside $r_{crit}$, and can be similarly estimated using
\citep{1976MNRAS.176..633F}\begin{equation}
\dot{n}_{lc}\simeq n(r_{crit})\,\Sigma_{BH}(r_{crit})\, v_{*}(r_{crit})\,\frac{\theta_{LC}^{2}(r_{crit})}{2}\label{eq:ndot-empty-lc}\end{equation}
where $\theta_{LC}^{2}=\frac{2r_{t}}{3r}$ is the loss-cone angle,$r_{crit}$
 the critical radius, and $r_{t}$ and $r_{w}$ are replaced with
$r_{crit}$ in equation (\ref{eq:cross-section-BH}). The term $\theta_{lc}^{2}/2$
represents the fraction of stars with velocity vectors pointing into
the loss cone for an isotropic velocity distribution.    Using the
same substitutions as above, equation (\ref{eq:ndot-empty-lc}) can
be written as  \begin{equation}
\dot{n}_{lc}\simeq8\sqrt{\pi}\, n_{c}\, v_{c}\, r_{i}\,\left(\frac{r_{i}}{r_{crit}}\right)^{5/4}r_{t}.\label{eq:ndot-loss-cone}\end{equation}
From the ratio\begin{equation}
\frac{\dot{n_{w}}}{\dot{n}_{lc}}\simeq\frac{12}{8}\left(\frac{r_{crit}}{r_{w}}\right)^{5/4}\label{eq:ratio-ndot-lc-w}\end{equation}
as well as from equation (\ref{eq:ndot-wandering}) we see, as we
expected, that, in general, the disruption rate is decreasing with
increasing wandering radius. Furthermore, when we calculate this ratio
 using $r_{w}\approx4\times10^{-3}$,  and $r_{crit}\approx2\times10^{-3}$ from
Fig. 1 in \citep{2004ApJ...613.1133B} for a cluster with $M_{BH}\approx700\,{\rm M}_{\odot}$
and $N=8\times10^{4}$ (run BME-1)  we find a value  of $\approx0.6$
. This is very similar to what we obtain for the ratio of disruption
rates in BME-16 (see Fig. \ref{fig:disrupt-no-substeps}). 
However, equation (\ref{eq:ratio-ndot-lc-w})
does not seem to hold for larger IMBH masses. If we calculate the
ratio for, e.g., the cluster with $M_{BH}\approx1395$ and $N=8\times10^{4}$
in the same figure (run BME-2), equation (\ref{eq:ratio-ndot-lc-w})
predicts that the disruption rate for a fixed IMBH should be lower
than for a wandering IMBH by a factor of $\approx1.6$ while from
Fig. \ref{fig:disrupt-no-substeps} we find that it is mostly larger.
The reason for that is probably related to the assumption of a full
loss-cone when deriving equation (\ref{eq:ndot-wandering}), as it
is clear that in the limit of very small $r_{w}$, and consequently
larger IMBH masses, the empty loss-cone regime, and thus, equation
(\ref{eq:ndot-loss-cone}), must be approached. Indeed, in Fig. \ref{fig:disrupt-no-substeps}
(left panel) we see that the disruption rates seem to converge at
late times. Similarly, we find very good agreement in the total number
of disruptions between $N$-body and our Monte Carlo runs for IMBHs
with even larger masses. As $r_{w}\sim\sqrt{M_{*}/M_{BH}}$ \citep{2002ApJ...572..371C},
it appears that the effect of a wandering IMBH has a negligible
influence on the disruption rate for $M_{BH}/M_{*}\gtrsim1000$.
Given that in the core of a multi-mass cluster with a central IMBH 
$M_{*}\approx 0.6\,{\rm M_\odot}$
\citep{2004ApJ...613.1143B}, we would expect that an IMBH can be treated 
as being fixed at the cluster center as long as 
$M_{BH}\gtrsim 600\,M_{\odot}$.

A third effect that could decrease the disruption rate and that is
not included in the current Monte Carlo scheme is the formation of,
and the strong interaction with, an IMBH binary that is able to scatter
other stars into the outer cluster regions. This mechanism has been
recently suggested by \citet{2008ApJ...686..303G} to suppress mass-segregation
in a multi-mass cluster with IMBH. \citet{2008ApJ...686..303G} also
found that the efficiency of this scattering process is relatively
independent of the relative mass of the IMBH. Given the trend we see
for the difference in the number of tidal disruptions between our
simulations and $N$-body results as a function of $M_{BH}/M_{c}$,
it appears that this mechanism is unlikely to be the main reason.

\begin{figure}
\includegraphics[width=0.8\paperwidth]{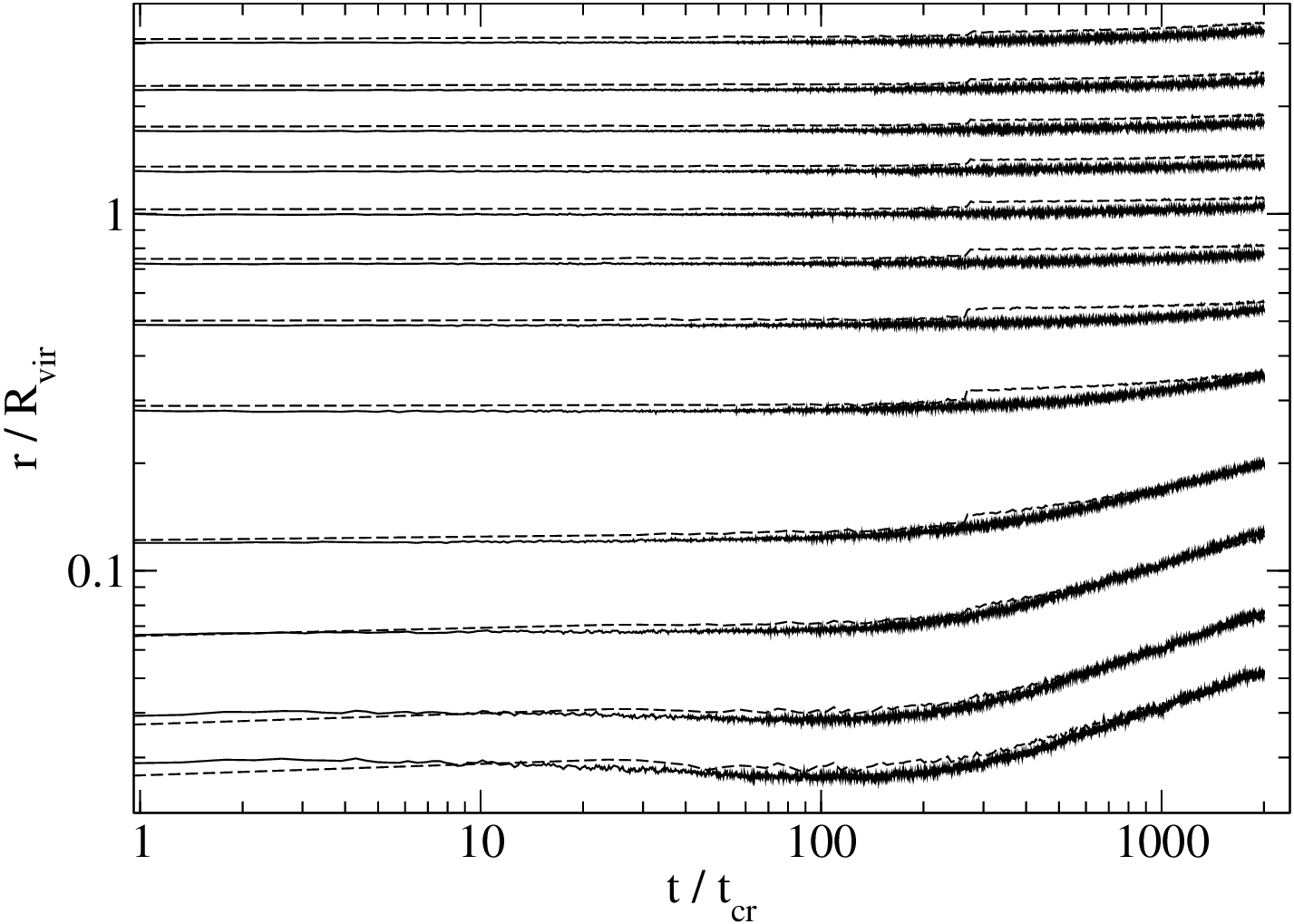}

\caption{\label{fig:Evolution-Lagrad}Evolution of the Lagrange radii of run
BME-16 for our Monte Carlo simulations (solid lines) and direct $N$-body
simulations (dashed lines). Shown are the radii that contain $1\%$,
$2\%$, $5\%$, $10\%$, $20\%$, $30\%$, $40\%$, $50\%$, $60\%$,
$70\%$, $80\%$, and $90\%$ of the total mass of the cluster. The
cluster expands due to the energy generated in the cluster center
by the tidal disruption of stars. Apart from a small jump in the $N$-body
data for the outer $90\%$ of the cluster, which appears to be spurious,
our results are in reasonable agreement with the results of \citet{2004ApJ...613.1133B}. }

\end{figure}

Finally there is the possibility that, as already argued in \S\ref{sub:Comparison-with-{Amaro-Seoane}},
the larger disruption rates are a result of the slower expansion of
the inner regions caused by the larger timestep size relative to the
local relaxation time. However, while this slower expansion might
increase the disruption rate in general, it is not at all clear why
this should make a larger difference for clusters with lower-mass
IMBHs. For instance, when we compare the ratios of the final masses
from our runs to the results of the individual time step code of \citet{2002A&A...394..345F}
for the different $M_{BH,i}$ we see from Fig. \ref{fig:AS01} that
they are rather similar ($0.70$ for $M_{BH,i}=50\, M_{\odot}$ and
$0.71$ for $M_{BH,i}=500\, M_{\odot}$). On the other hand, the final
mass ratios for runs BME-1 to BME-3 vary dramatically in comparison
(from $0.6$ to $0.95$) for a similar range in initial IMBH masses.
Another reason why it is less likely that the differences to the direct
$N$-body results are caused by differences in the expansion rates
is given in Fig. \ref{fig:Evolution-Lagrad}, where we compare the
Lagrange radii for run BME-16. As can be seen, apart from very minor
deviations, the expansion of the cluster in our simulation agrees
well with the $N$-body results, even for the regions within $0.1\, r_{h}$.
However, one should also note that, here, only the radii are shown
that contain more than $1\%$ of the total cluster mass, as only for
those $N$-body data were available. Larger differences would be expected
for smaller radii, especially because for this particular run the
$1\%$ Lagrange radius is just outside the cusp region which mostly
determines the disruption rate.

In general we can say that, despite the significant differences in
the total number of disrupted stars, the disruption rates do not differ
greatly from the ones from direct $N$-body simulations and are even
in very good agreement for runs with $M_{BH}/M_{c}\gtrsim0.01$ given
the error bars. Furthermore, from Fig. \ref{fig:Evolution-Lagrad}
we see that our Monte Carlo code can reproduce the evolution of the
cluster structure, at least down to the cusp region, quite well. We
expect that this also applies to other clusters as long as the wandering
radius of the IMBH is inside the cusp.

\section{Realistic Cluster Models\label{sec:Realistic-Cluster-Models}}

\subsection{Comparison to BMH}\label{Comparison-to-BMH}

Here we re-examine a subset of the $N$-body models of BMH using our
Monte Carlo code to see if we are able to reproduce their results
and, in particular, the inner surface brightness slopes mentioned
in \S\ref{sec:Previous-Work-of}. For this comparison we only include
runs with large $N$ in order to populate the high mass end of the
IMF sufficiently well, which is important for the applicability of
the Monte Carlo method. Furthermore, as in BMH, we only include bright
stars, defined as main-sequence stars and giants with masses larger
than $90\%$ of the turn-off mass, for the calculation of the surface
brightness profiles to take into account that those stars contribute
most ($>80\%$) of the observed light. 

Table \ref{tab:mc-bmh} gives a summary of the resulting cluster structural
parameter after the clusters have been evolved to an age of $12\,{\rm Gyr}$
together with the corresponding results of BMH. In contrast to our
single-mass runs we find lower IMBH masses in our runs compared to
the direct $N$-body results. However, this difference is much less
pronounced, if not negligible, as it is at most $20\,{\rm M}_{\odot}$,
resulting in final IMBH masses that deviate at most by $7\%$ from
the $N$-body results. The reason for that is not only because the
tidal disruption radii of the stars are generally lower compared to
our single-mass cases but, more importantly, the cluster had a much
lower density in the core initially and subsequently expanded because
of mass loss due to stellar evolution. After the core started contracting
again, after about $1\,{\rm Gyr}$, it became quickly dominated by
massive dark remnants, mostly black holes and massive white dwarfs,
that drove out lower-mass main-sequence stars. This decreases the
overall disruption rate, because the main-sequence stars which are
more easily disrupted given their much larger radii compared to compact
remnants, are, on average, much further out, while the compact remnants,
though much closer to the IMBH, are unlikely to get disrupted.

The fact that for these simulations there are fewer disruptions in
our Monte Carlo runs than in the $N$-body simulations is most probably
again related to the wandering of the IMBH. While for a fixed IMBH
there is a well populated cusp, for a wandering IMBH no such clear
cusp can be identified. In the $N$-body simulations the massive dark
remnants in the cusp are ejected through strong gravitational interactions
with the IMBH, which allows the main-sequence stars and giants to
diffuse inwards and to come closer to the tidal radius. In addition,
due to its motion, the IMBH is also able to come closer to stars that
are just outside of the cusp region, which also increases the number
of disruptions. 

Comparing the cluster structural parameters we find again very good
agreement between the Monte Carlo and direct $N$-body runs. There
are only minor but systematic differences inasmuch as in our Monte
Carlo runs the clusters have systematically larger masses, though
only by less than $1.5\%$, and are more compact, with their half-mass
radii differing at most by $10\%$ . The most likely reason is that,
in our simulations, we do not model close encounters with stars tightly
bound to the IMBH. Those interactions might frequently lead to the
ejection of massive stars or remnants, which also contributes to the
cluster expansion. In addition, the somewhat larger disruption rates
might have also added to the stronger expansion. However, given the
minor differences between the Monte Carlo and $N$-body results, close
encounters with stars tightly bound to the IMBH do not seem to have
a significant influence on the cluster structure as a whole.

\begin{figure}
\begin{tabular}{cc}
\includegraphics[clip,width=0.5\textwidth]{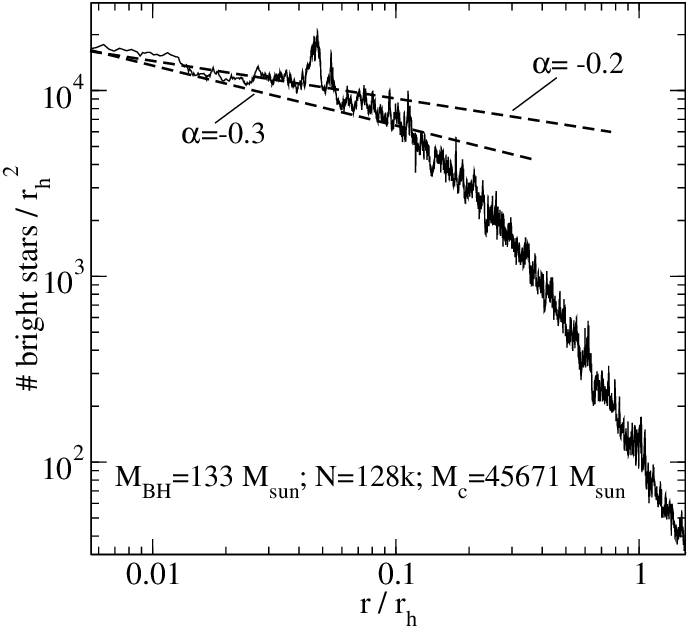}  & \includegraphics[clip,width=0.5\textwidth]{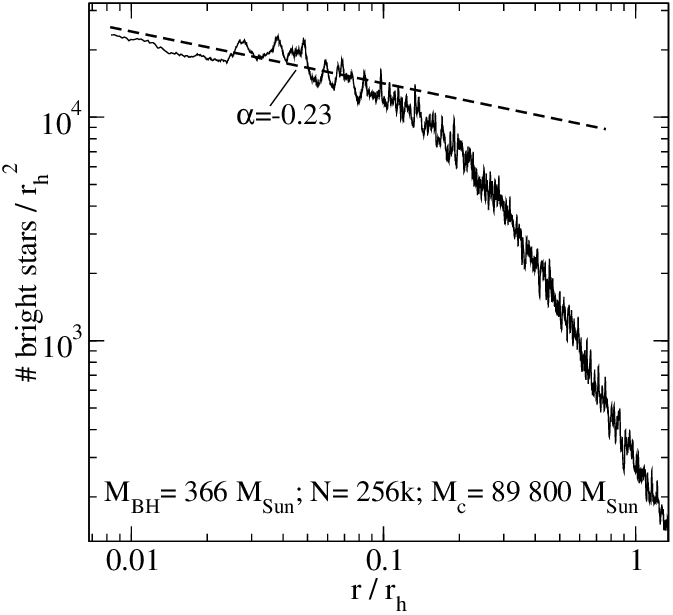}\tabularnewline
\end{tabular}

\caption{Surface density profile of bright stars for two clusters with different
numbers of stars and BH masses. The dashed line in the right panel
is a power-law fit to the inner region of the cluster, while the two
dashed lines in the left panel are power-laws with slopes bracketing
the range $[-0.2,-0.3]$ suggested by BMH for clusters harboring IMBHs.
The inner parts of our surface density profiles are in good agreement
with the results of BMH.}

\label{surfaceDensity} 
\end{figure}

We also achieve good agreement for the surface density profiles between
the two methods. Fig.\ref{surfaceDensity} %
 shows the two-dimensional density profiles of bright stars for two
clusters at the end of the simulation. These profiles are obtained
by averaging five snapshots obtained at $50\,{\rm Myr}$ intervals.
The profiles show only very shallow cusps with a power-law slopes
$\alpha$ between $-0.2$ and $-0.3$, consistent with the $N$-body
results. In order to reduce the amount of noise and obtain a more
reliable fit of the inner slope of the surface brightness profile,
we re-ran run BMH-2 but with twice as many particles. The resulting
profile is shown in Fig. \ref{surfaceDensity}. As one can see, the
power-law fit has a slope of $\alpha=-0.23$, very close to the average
$\alpha=-0.25$ found in \citet{2005ApJ...620..238B} and agreeing
very well with the relation (\ref{eq:miocci}) found by \citet{2007MNRAS.381..103M}.

\subsection{Comparison to Real Star Clusters\label{sub:Comparison-to-Real}}

The ultimate goal of any cluster simulation is to reproduce observations
of real star clusters as closely as possible. For globular clusters
these observations are mostly in the form of photometric data and
surface brightness profiles, while there are only a few clusters for
which well measured velocity dispersion profiles are available. In
the previous section we found that our Monte Carlo simulations with
IMBH as well as the corresponding $N$-body simulations of \citet{2005ApJ...620..238B}
are able to reproduce the intermediate inner surface brightness slopes
seen in some Galactic globular clusters. As discussed in 
\S\ref{sec:Previous-Work-of}, an IMBH also influences the surface brightness
profile in that it produces rather large cluster cores as measured by
$r_{c}/r_{h}$.  It is, therefore, interesting to see if the combination of
slope and concentration we find in our models matches any observations.
However, instead of comparing slopes and concentrations quoted in the
literature it is more suitable to directly compare surface brightness profiles
in this case, given that the tidal radii of observed clusters are very
uncertain \citep[see, e.g., discussion in][]{2010MNRAS.401.1832B}. 

For this comparison we choose the cluster NGC 5694 for which
\citet{2006AJ....132..447N} report an inner slope of $0.19\pm0.11$, close to
our fit in Fig. \ref{surfaceDensity} (right panel). We modeled the cluster
by carrying out a large parameter survey consisting of approximately 600 model
calculations, varying initial number of stars, concentration, virial radii, and
IMBH mass. All models had the same IMF as in BMH and were evolved for 12 Gyr.
As the orbit of NGC5694 in the Galaxy is not known we assumed for simplicity
that it moves on a circular orbit at its current distance of $\approx29\,{\rm
kpc}$ from the Galactic center. Given this distance, it is a rather isolated
cluster, and has been speculated to be of extra-galactic origin
\citep{2006ApJ...646L.119L}. We included in our search also clusters that
contain no IMBH but, instead, $10\%$ hard binaries, as it has been shown that a
cluster with binaries can possess similar shallow surface brightness slopes as
a cluster harboring an IMBH \citep{VesperiniTrenti}. We calculated the surface
brightness profile by converting the stellar radius and bolometric luminosity
for each star obtained from the CMC stellar evolution module BSE to V-band
luminosity using the standard stellar library in \citet{1998A&AS..130...65L}.
The luminosities were then radially binned similar to
\citet{2006AJ....132..447N} and converted to apparent magnitudes using a
distance of 35~kpc from the sun. In order to minimize the large fluctuations
caused by the brightest giants, we only consider objects with an absolute
V-band magnitude fainter than 3. This value is also low enough to ensure that
the shape of the surface brightness profile does not become significantly
biased \citep[see also][]{GierszHeggie2009a}. Table \ref{NGC5694-Models}
summarizes the parameter ranges explored.

\begin{table}[p]
\begin{tabular}{cc}
Parameter & Range\tabularnewline
\hline
$N$ & $0.6-1.7\times10^{6}$\tabularnewline
$r_{vir}$ & $2-4\,{\rm pc}$\tabularnewline
$W_{0}$ & $0.8-7$\tabularnewline
$M_{BH}$ & $500-4500\,{\rm M_{\odot}}$\tabularnewline
$R_{G}$ & $29.4\,{\rm kpc}$\tabularnewline
$Z$ & $0.0004$\tabularnewline
\end{tabular}

\caption{\label{NGC5694-Models} Parameter ranges explored. The galactic
distance, $R_{G}$, and metallicity,
$Z$, are from the Harris catalog (Harris 1996).}
\end{table}

\subsubsection{Surface Brightness Profiles}

\begin{figure}[p]
\begin{tabular}{cc}
\includegraphics[clip,width=0.5\textwidth]{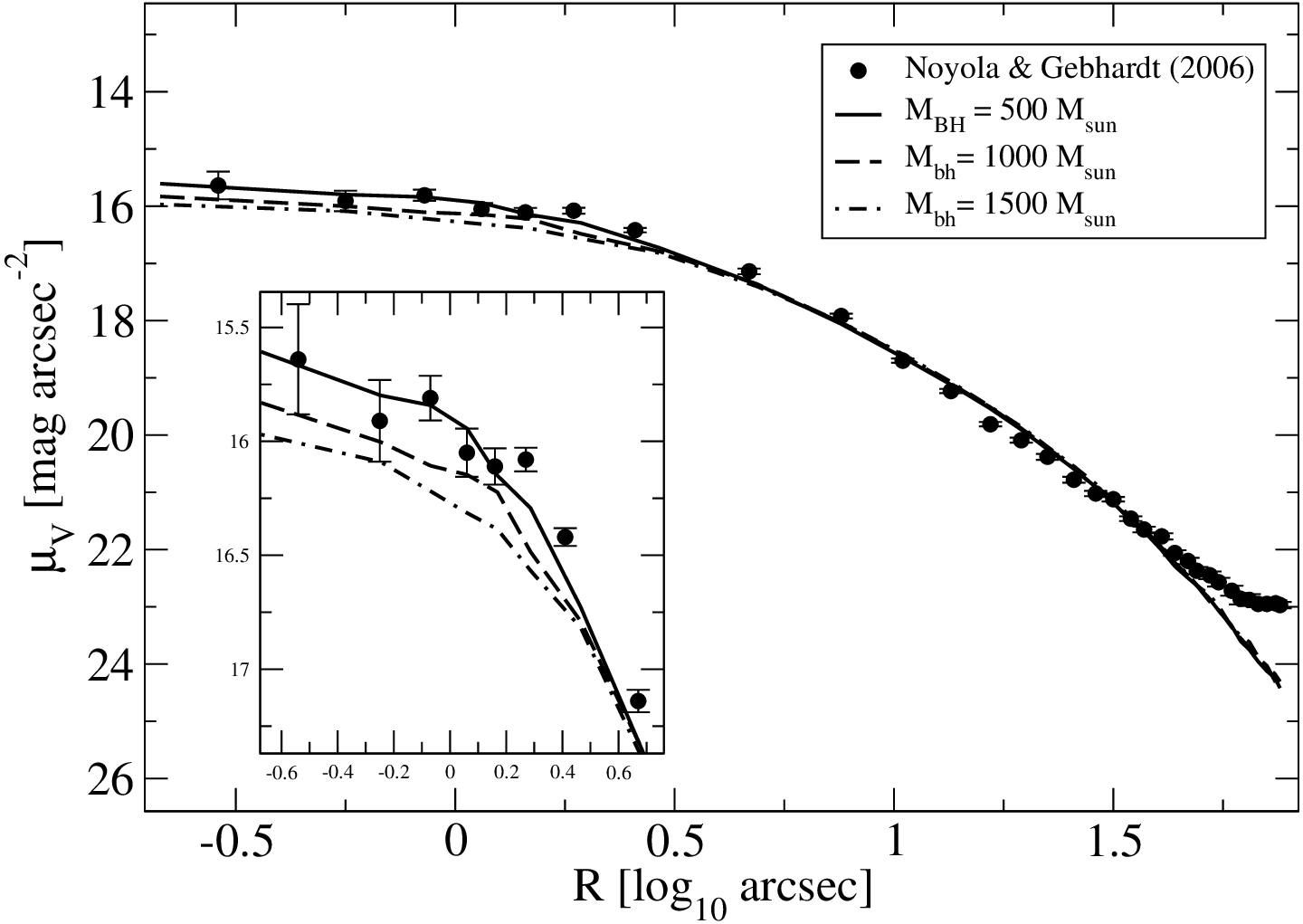}  & \includegraphics[clip,width=0.5\textwidth]{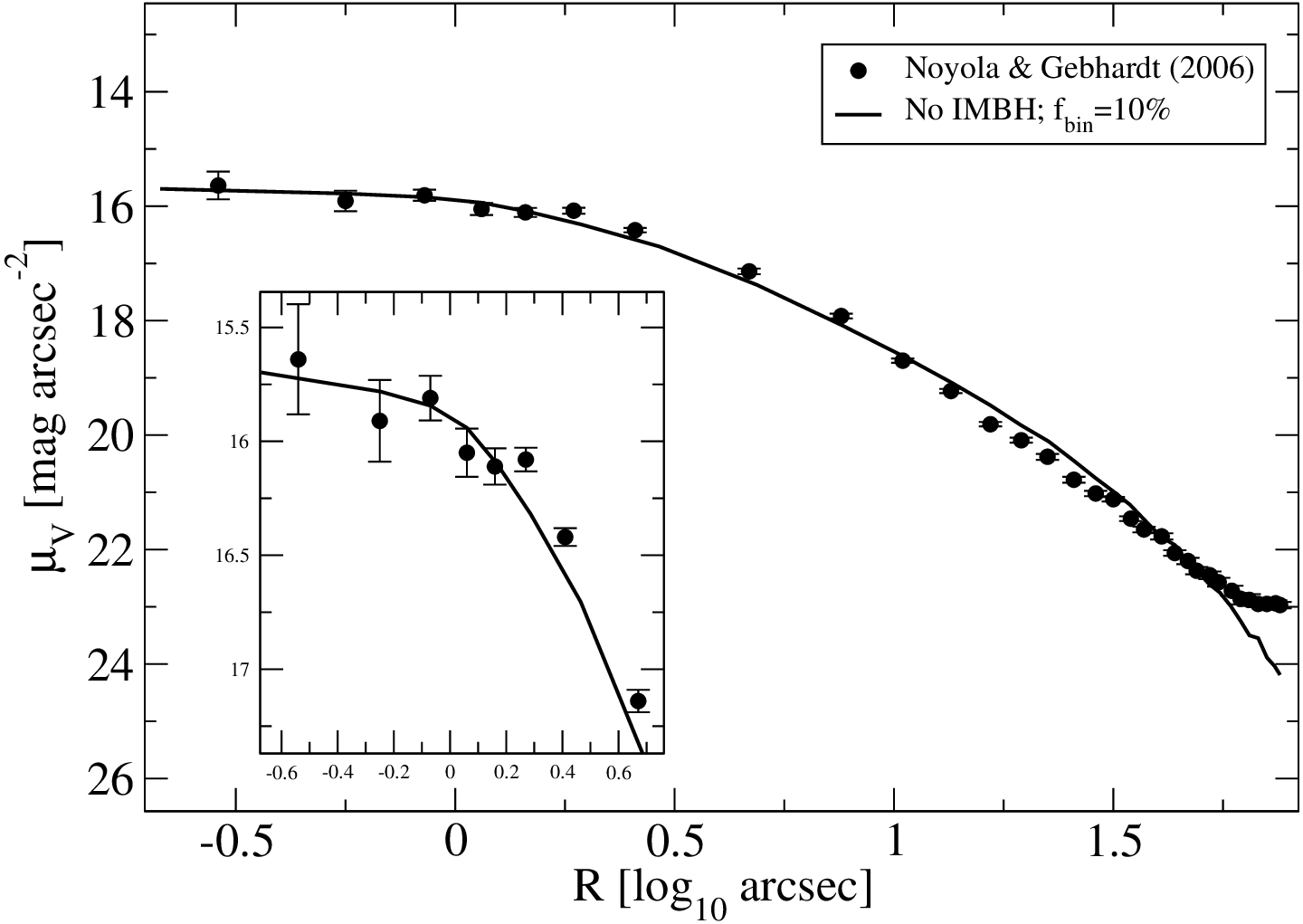}\tabularnewline
\end{tabular}

\caption{\label{fig:Surface-brightness-profile}Surface brightness profile (SBP)
from our best fit model with (left panel) and without (right panel) central
IMBH for NGC 5694 (Noyola \& Gebhardt 2006; filled circles), a cluster that
might harbor an IMBH \citep{2005ApJ...620..238B}, showing a shallow cusp.
Right: Models with different IMBH masses. The maximum IMBH mass for which a
reasonable match to the data can be obtained is $1000\,{\rm M}_\odot$.}
\end{figure}

Fig. \ref{fig:Surface-brightness-profile}%
 shows the resulting surface brightness profiles of our best-fit
models along with the data from \citet{2006AJ....132..447N}, and in Table
\ref{best-nobh} and \ref{best-bh} are their initial and final cluster
parameters. 

\begin{table}[p]
\begin{tabular}{ccc}
 & $t=0$ & $t=12\,{\rm Gyr}$\tabularnewline
\hline
$N$ & $1.30\times10^{6}$ & $1.28\times10^{6}$\tabularnewline
$M_{c}$ & $8.08\times10^{5}\,{\rm M_{\odot}}$ & $4.83\times10^{5}\,{\rm M_{\odot}}$\tabularnewline
$r_{h}$ & $2.1\,{\rm pc}$ & $4.4\,{\rm pc}$\tabularnewline
$t_{rh}$ & $0.96\,{\rm Gyr}$ & $3.75\,{\rm Gyr}$\tabularnewline
$r_{tide}$ & $308\,{\rm pc}$ & $260\,{\rm pc}$\tabularnewline
$f_{bin}$ & $10\%$ & $8.9\%$\tabularnewline
\end{tabular}

\caption{\label{best-nobh}Evolution of the characteristics of our best-fit
model without IMBH. Here, $N$ is the number of stars and binaries, $r_{tide}$
the tidal, or Jacobi, radius, and $f_{bin}$ the binary fraction. The initial
density distribution is a King profile with $W_{0}=3$.}
\end{table}

\begin{table}[p]
\begin{tabular}{ccc}
 & $t=0$ & $t=12\,{\rm Gyr}$\tabularnewline
\hline
$N$ & $1.30\times10^{6}$ & $1.28\times10^{6}$\tabularnewline
$M_{c}$ & $7.63\times10^{5}\,{\rm M_{\odot}}$ & $4.65\times10^{5}\,{\rm M_{\odot}}$\tabularnewline
$r_{h}$ & $1.98\,{\rm pc}$ & $3.98\,{\rm pc}$\tabularnewline
$t_{rh}$ & $0.90\,{\rm Gyr}$ & $3.29\,{\rm Gyr}$\tabularnewline
$r_{tide}$ & $308\,{\rm pc}$ & $261\,{\rm pc}$\tabularnewline
\end{tabular}

\caption{\label{best-bh}Evolution of the characteristics of our best-fit model with a
$500\,{\rm M_{\odot}}$ IMBH. The initial density distribution is a King
profile with $W_{0}=0.8$.}
\end{table}

In general, the model surface brightness profiles match well within $60\,{\rm
arcsec}$, while outside, the observed profile seems to flatten.  This excess
light is usually attributed to both, the confusion with background stars and
stars that are escaping but remain still close to the cluster for a
considerable time \citep{2000MNRAS.318..753F}, which we cannot model using a
simple tidal cut-off prescription. In this case, however, confusion with
background stars is a more likely explanation as the estimated tidal radius
from the literature \citep{1996AJ....112.1487H} and the Jacobi radius we obtain
at the end of our models is larger by a factor of $4$ compared to the radius
where the observed profile begins to flatten. 

Comparing the inner profile we find that, indeed, within the photometric
errors, the data can be well reproduced even without IMBH, while models with
IMBH show good agreement if the IMBH mass is less than $\approx1000\,{\rm
M}_{\odot}$. As the latter mass corresponds to a BH-to-cluster mass-ratio of
$0.2\%$, with a total cluster mass of $4.7\times 10^{5}\, {\rm M}_\odot$ at the
end, our best-fit IMBH masses are in good agreement with the extrapolated
relation for SMBH-harboring galaxy bulges.

The good agreement of the surface brightness profiles for the cluster model
with IMBH is somewhat surprising as, according to \citet{2007MNRAS.381..103M},
NGC 5694 is not expected to harbor an IMBH based on the fact that its
concentration is too large for the relatively steep inner surface brightness
slope (see relation \ref{eq:miocci}).  However, this statement is based on
values for the concentration quoted from the literature
\citep{1996AJ....112.1487H} which, as we mentioned before, are in general
rather uncertain due to difficulties determining the tidal radius of a cluster.
An error of $40\%$ in the measured tidal radius would bring the value of the
slope and the concentration of the cluster in agreement with relation
(\ref{eq:miocci}). Furthermore, the observed slope has a rather large
uncertainty as well, which, even considered alone, could make the observed
parameters consistent with this relation.  Therefore, it appears that the close
match we obtain for the shape of the surface brightness profile in our
simulation with the observed one of NGC 5694 does not necessarily contradict
the validity of the slope-concentration relation found by
\citet{2007MNRAS.381..103M} but seems to be broadly consistent with it.

\subsubsection{Time Variability}

Since \citet{GierszHeggie2009a} and \citet{GierszHeggie2009b} the importance of
fluctuations when comparing cluster models with observations has been better
understood. More recently \citet{VesperiniTrenti} investigated the variation of
the inner surface brightness slopes for clusters with and without IMBH based on
$N$-body models with up to $65536$ stars, and found that intermediate slopes in
the range expected for clusters harboring an IMBH are also ubiquitous among
clusters without IMBH, in particular when there is a non-negligible fraction of
dynamically hard binaries. However, the models in \citet{VesperiniTrenti} were,
owing to the computational expense of direct $N$-body simulations including
binaries, rather idealized in the sense that they only contained a much lower
number of particles, by a factor of $20$, than are actually present in NGC5694.
In addition, the surface brightness slopes were derived by considering all
main-sequence stars while most of the light is actually contributed by only a
small subset of main-sequence stars and giants with masses larger than
$\approx0.7\,{\rm M_\odot}$ for clusters with ages of around $12\,{\rm Gyr}$
(see also BMH and \S \ref{Comparison-to-BMH}). It is likely that these
limitations will influence the derived surface brightness slopes.

\citet{2011ApJ...743...52N} raised similar concerns. They analyzed a series of
low-$N$, $N$-body models with and without IMBHs, that were stacked on top of
each other to contain in the end up to $6\times 10^6$ stars for constructing
corresponding HST-like images and then deriving surface brightness slopes the
same way as has been done for observed GCs in \citet{2006AJ....132..447N}.
Contrary to \citet{VesperiniTrenti}, their results show that the inner surface
brightness slope is a good diagnostic to discern IMBH-less clusters from likely
candidates of IMBH harboring clusters, with intermediate values, ranging from
-0.1 to -0.4, indicating the presence of an IMBH.

Given these discrepancies and limitations in the literature, it is worthwhile
to investigate the slope variations for our full-scale models, where the
cluster as well as single and binary evolution is self-consistently taken into
account. Although we have not checked whether our Monte Carlo code is able to
reproduce the density fluctuations of an $N$-body model,
\citet{GierszHeggie2009b} show that the time variations of the core radius in
their Monte Carlo model, which they identify as to be due to variations of the
inner density slope, are of the same magnitude as in a corresponding $N$-body
model, though less coherent due to the random sampling of the stellar orbits.
It thus appears that the density fluctuations in a Henon-type Monte Carlo
code reflect the degree of variations present in $N$-body simulations and can
at least be used to get some first insight into the temporal behavior of the
surface brightness profile. Ideally, we would like to follow the inner surface
brightness profile with a direct $N$-body simulation for a brief time period
around the current age of NGC5694, but even for such a relatively short time a
direct $N$-body calculation of the entire cluster with $N\approx 1.3\times
10^6$ stars would take a prohibitive amount of time.

For this analysis, we used a simple least-square fit for the determination of
the SBP slope over a range that covers the innermost 3 points of the
\citet{2006AJ....132..447N} profile.  This range corresponds to $11$ to $32\%$ of
the core radius, and is, thus, similar to the one \citet{VesperiniTrenti} used
for their determination of slopes. In order to reduce the uncertainty of the
individual fits we used a much finer binning than in Fig.
\ref{fig:Surface-brightness-profile}, with 50 logarithmically spaced bins in
the fitted range.

\begin{figure}[p] \includegraphics[width=0.8\paperwidth]{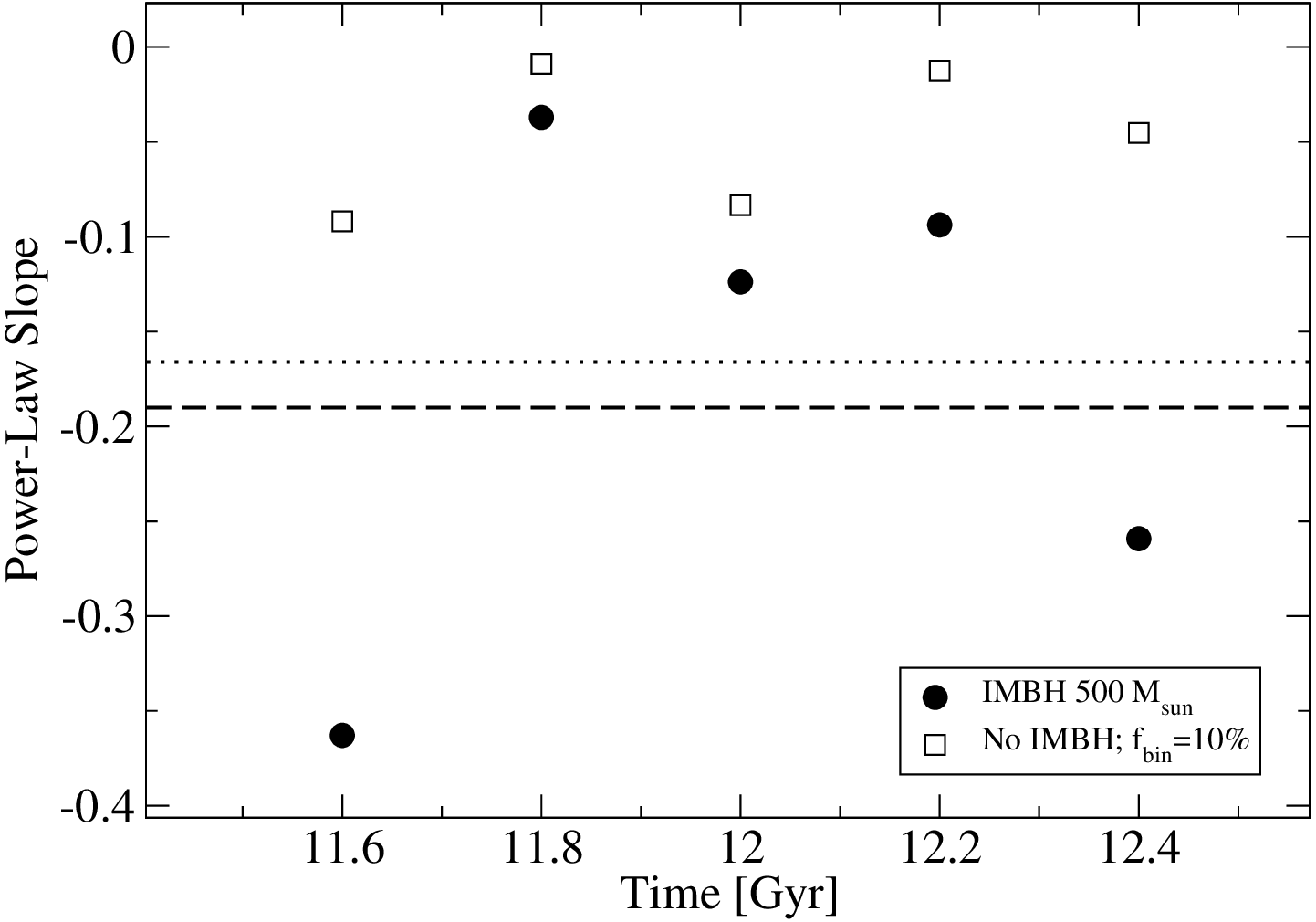}
\caption{\label{slopes-over-time} Evolution of the inner surface brightness
slopes for our best fit models with (filled circles) and without (open squares)
IMBH. The dashed line is the slope value as measured by \citet{2006AJ....132..447N}
($-0.19$), while the dotted line is the value from a least-square fit to their
three innermost data points ($-0.17$), which covers the range between $0.28$
and $0.85\,{\rm arcsec}$, and roughly corresponds to 11 to 33\% of the core
radius. The slopes in the model without IMBH always remain significantly
shallower ($>-0.1\%$) than the observed one, while in the models with IMBH they
cover a much wider range, from $>-0.1$ to $-0.4$. The uncertainties of the
fitted slopes are always $\approx0.1$} \end{figure}

\begin{figure}[p] \includegraphics[width=0.8\paperwidth]{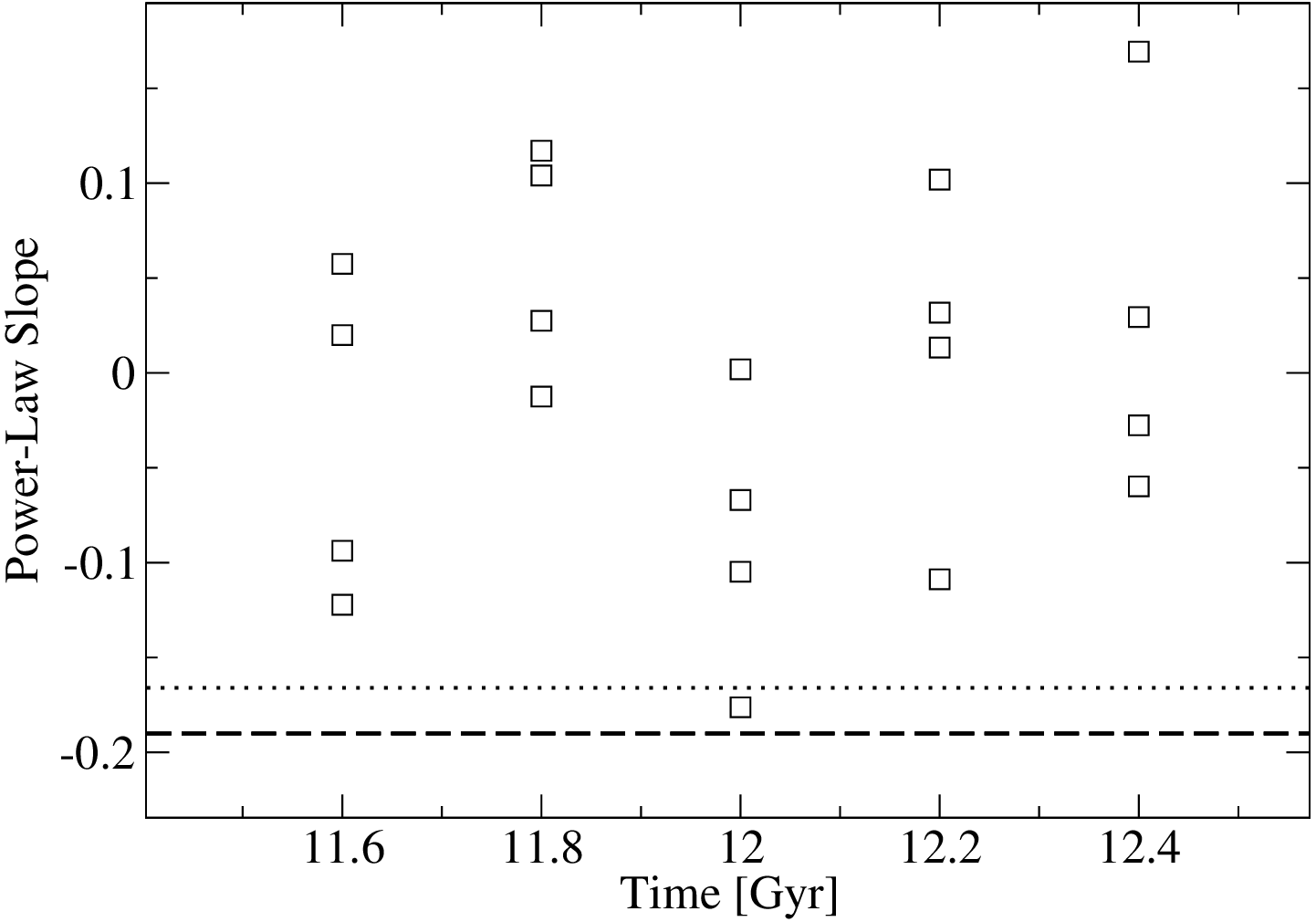}
\caption{\label{slopes-over-time-nobh} Evolution of the inner surface
brightness slopes for models similar to the best fit model without IMBH but
slightly lower concentrations (open squares). The dashed line is the slope
value as measured by \citet{2006AJ....132..447N} ($-0.19$), while the dotted line is
the value from a least-square fit to their three innermost data points
($-0.17$), which covers the range between $0.28$ and $0.85\,{\rm arcsec}$, and
roughly corresponds to 11 to 33\% of the core radius. With the exception of one
point, the fitted slopes never exceed $-0.17$, the value of the least-square
fit to the data. The uncertainties of the fitted slopes are always
$\approx0.1$} \end{figure}

Fig. \ref{slopes-over-time} shows the evolution of the inner surface brightness
slopes from 5 snapshots covering a period of $1\,{\rm Gyr}$ at around $12\,{\rm
Gyr}$. As can be seen, the slopes of the cluster models without IMBH do not
steepen much beyond $-0.1$, while for the cluster with IMBH they cover a much
wider range, from $>-0.1$ to $-0.4$. Comparing with the slopes derived from
observations it appears, at face value, somewhat unlikely that the cluster
model without IMBH can match the observed slope, and a cluster with IMBH seems
more likely to fit. On the other hand, the uncertainties of the individual
slopes are, with $0.1$, rather large, so the disagreement is only marginally
statistically significant. In addition, it might pretty well be that due to our
coarse sampling we have missed possible maxima. For this reason we also
analyzed 4 similar models in our grid, with the same initial particle number
and radius but different, slightly lower concentrations, going down to
$W_0=1.8$ in steps of $0.2$ (Fig. \ref{slopes-over-time-nobh}). As one can see,
with the exception of one point, all slopes remain shallower than $-0.17$, the
value from our least-square fit to the observations. Although about a third of
the values are within the error bars, it is nevertheless remarkable that the
average slope is close to zero while for models with IMBH it is clearly steeper
than $-0.1$ and closer to the observed one. Thus, based on our results it
remains, at least on a qualitative level, somewhat more difficult to reconcile
a cluster model without IMBH with observations.

Clearly, given the large uncertainties of not only our slopes, but also of the
slope in \citet{2006AJ....132..447N} for NGC5694, a more detailed statistical
analysis, taking the surface brightness profile directly into account, is
certainly desirable in order to quantify how likely or unlikely it is that
NGC5694 harbors an IMBH. As \citet{GierszHeggie2009a} already pointed out, the
agreement between model and observations has to determined in terms of the
probability that an observational profile be rejected as a member of the
ensemble of profiles provided by the model. Such an analysis, however, is
beyond the scope of the present paper.

\subsubsection{Dynamical Age}

The advantage of having modeled the evolution of NGC 5694 self-consistently
with a realistic number of stars is that it allows us to obtain a more reliable
estimate of its dynamical age. The dynamical age has consequences for the
expected inner surface brightness slope for an observed cluster
\citep[e.g.,][]{2006AJ....132..447N,Trenti10}, and is essential when assessing
possible mass-segregation signatures for IMBHs \citep{2008ApJ...686..303G}.

\citet{Hurley} already emphasized that the relaxation time of a cluster is
time-dependent and the actual dynamical cluster state, whether in the
core-contraction or past core-collapse phase, is better described in terms of
$N_{t_{rh}}$, the number of elapsed half-mass relaxation times, defined as
\begin{equation}N_{t_{rh}}=\int\limits_{0}^{\tau}\frac{dt}{t_{rh}(t)}\,,
\label{Ntrh} \end{equation} where $\tau$ is the cluster age. He showed that in
his models $t_{rh}$ decreases by a factor of two over the cluster lifetime and
compared $N_{t_{rh}}$ with simple estimates, either dividing the cluster age by
the initial or current $t_{rh}$, finding that the former estimate was much
closer to $N_{t_{rh}}$ than the latter.

Apparently assuming the same behavior for the evolution of NGC 5694 and using
an estimate for the current half-light relaxation time of $t_{rh}=1.9\,{\rm
Gyr}$ from the updated Harris catalog \citep{1996AJ....112.1487H},
\citet{VesperiniTrenti} estimated its dynamical age as $N_{t_{rh}}\approx 3$
(see their Fig. 3).  However, they also point out that such estimated ages have
rather large uncertainties, and detailed models are necessary in order to get
more accurate results.

In fact, as inspection of Table \ref{best-nobh} and \ref{best-bh} reveals, the
simple extrapolation of the \citet{Hurley} results to our full-scale models is
not justified. This is mainly because NGC 5694 is not tidally limited and can
expand freely which increases its relaxation time rather than decreasing it
over time. Calculating the dynamical age of our models for NGC 5694 as in eq.
\ref{Ntrh}, we find $N_{t_{rh}}= 4.2$ for the best-fit model without and
$N_{t_{rh}}= 4.7$ for the one with IMBH. Therefore, NGC 5694 appears to be not
far from core collapse and has evolved considerably.  Straightforward estimates
based on the initial and final relaxation time, on the other hand, give
$\tau/t_{rh}(0)\approx13$ and $\tau/t_{rh}(\tau)\approx3$ respectively, which
means that, contrary to the case in \citet{Hurley}, using the current instead
of the initial half-mass relaxation time gives a better approximation for the
dynamical age of this cluster. Considering, however, that even this estimate
differs by as much as $40\%$ and the difference still worsens when the cluster
gets older and expands further, neither approximation seems really suitable for
a reliable estimate.

Coincidentally, if the value quoted from the Harris catalog is corrected for
the fact that the projected half-mass radius is smaller by $\approx25\%$ than
the unprojected one \citep[see, e.g.,][]{Hurley}, the resulting current
$t_{rh}\approx2.8\,{\rm Gyr}$ would imply $N_{t_{rh}}=4.4$, which is very close
to the integrated value from Equation \ref{Ntrh} for our models. Clearly, one
cannot attach any meaning to this as the corrected relaxation time still
differs significantly from the one in our model, owing to the underlying
simplifying assumptions \citep[see][]{1996AJ....112.1487H}, and the close match
is rather accidental.

As this comparison shows, there seems to be no simple way to reliably estimate
dynamical ages of clusters without considering the previous cluster evolution
in more detail.

\section{Summary and Conclusions\label{sec:Summary-and-Conclusions}}

In this paper we studied the influence of IMBHs on the evolution of
globular clusters using our Monte Carlo code, which has been extended
to include the effects of a central IMBH on the stellar distribution.
The IMBH is treated as a fixed point mass in the cluster center, and
at each time step stars are tested for entry into their loss-cone.
 In order to test our implementation we carried out a large number
of idealized, as well as more realistic dynamical simulations of globular
clusters and compared our results with published results from direct
 $N$-body and Fokker-Planck calculations. 

In general we found that our results agree reasonably well with results
from $N$-body and Fokker-Planck type codes. Significant differences
only exist for idealized models while for more realistic clusters
these differences become negligible.

 We found that our code is able to reproduce the expected density
cusp and Keplerian velocity profile around the IMBH for a single-mass
cluster very well. In addition, we also found that the evolution of
the cluster density profile outside of the $1\%$ Lagrange radius
is in very good agreement with direct $N$-body simulation. The only
notable difference is that the cusp extends further down towards the
IMBH than in the direct $N$-body simulations of \citet{2004ApJ...613.1133B}.
The reason is most likely  that the IMBH is fixed at the cluster center
while in the $N$-body simulations it is allowed to move freely. Although
this has a minor effect on the density profile, causing it mainly
to be flatter inside $r_{w}$, we argue that it produces significant
differences in the disruption rates for single-mass clusters. In particular,
as we demonstrated by simple estimates of the disruption rate, the
effect of IMBH wandering might be able to explain why we see  up to
a factor of $2$ larger disruption rates for low-mass IMBHs than for
higher-mass ones, compared to $N$-body results, which is difficult
to explain otherwise. For a sufficiently fast moving IMBH the loss-cone
can always be assumed full in which case the disruption rate is a
strong function of the density at the disruption radius $r_{t}$.
Therefore, for larger $r_{w}$, and consequently for lower $M_{BH}$,
the density at $r_{t}$ becomes lower inside the very steep power-law
cusp which, in turn, decreases the disruption rate. In our calculations
 the disruption rate is determined by the diffusion of stars into
the loss-cone. This rate turns out to be larger than for the case
of a wandering IMBH for our cluster parameters if $r_{w}$ is significantly
larger than $r_{crit}$, which agrees with the differences we see
for run BME-16. For lower $r_{w}$, and, thus, larger $M_{BH}$, loss-cone
depletion effects appear to become important as we find from our simulations
that the rates for these two cases converge for IMBHs with $M_{BH}/M_{*}\gtrsim1000$,
while equation (\ref{eq:ratio-ndot-lc-w}) predicts that the rate
for a wandering IMBH should become larger than for a fixed one. 

The agreement for larger $M_{BH}$ also implies that, in a realistic GC, an IMBH
with $M_{BH}>600\,\rm{M_{\odot}}$ can be safely treated as fixed central point
mass, given the scaling of the wandering radius and the fact that the average
stellar mass in the core converges quickly to $M_{*}\approx
0.6\,\rm{M_{\odot}}$ \citep{2004ApJ...613.1143B} in multi-mass clusters with a
central IMBH. These larger IMBH masses are expected for clusters with
$M_c\gtrsim 2\times 10^5\, {\rm M_{\odot}}$ initially, based on a
straightforward extrapolation of the well known $M-\sigma$ relation
\citep{MagorrianRelation,GebhardtMsigma,FerrareseMerritt}. Since most GCs are
more massive than this, we conclude that our code is able to calculate the
evolution of GCs with a central IMBH for a wide range of relevant cluster
parameters.

Comparing the IMBH mass growth in the empty-loss-cone regime  with
different Fokker-Planck type codes, such as the gas code by
\citet{2004MNRAS.352..655A} and the Monte Carlo code by
\citet{2002A&A...394..345F}, we found better agreement with our results. For
all codes the IMBH growth shows the same qualitative behavior, that is, the
masses converge to a constant value.  Only at late times the masses differ
significantly by factors of 1.3 to 2. The larger IMBH masses in our simulations
compared to the results of \citet{2002A&A...394..345F} are most likely due to
the slower expansion of the inner regions in our simulations, which are caused
by the larger timesteps relative to the local relaxation time, a consequence of
the shared timestep scheme used in our code.  Contrary to the effect of IMBH
wandering, this effect does not introduce a dependence on $M_{BH}$ and our
final masses always differ by $30\%$ regardless of the initial $M_{BH}$. 

The situation for realistic cluster models is, however, rather different.
In this case the disruption rates are generally lower because the
cluster core gets quickly dominated by dark remnants, which have extremely
small $r_{t}$. Therefore, the influence of the IMBH wandering is,
in general, much reduced and does not produce significantly different
results compared to a fixed IMBH. There are only minor deviations
in the final cluster mass and size, with our clusters being more compact
by $\approx10\%$ and slightly more massive by $\approx1\%$. The
reason here might be related to the fact that we do not model close
encounters with stars in the cusp, which could be an important ejection
mechanism for the stars \citep{2004ApJ...613.1133B}. Such encounters
are also responsible to efficiently remove dark remnants from the
cusp \citep{2004ApJ...613.1143B} and, thus, allow for lower-mass
main-sequence stars to get close to the IMBH and disrupted. The somewhat
larger number of disruptions in the $N$-body simulations is also
caused by the wandering of the IMBH as it can, this way, get closer
to the main-sequence stars that are, due to mass segregation, further
out. Although, the total mass the IMBHs gain in the $N$-body simulations
is up to 3 times larger than in our simulations, the total difference
amounts to just $20\,{\rm M}_{\odot}$ at most or less than $7\%$
of the total cluster mass and is, thus, very minor, given that this
difference comes from not much more than 10 disrupted main-sequence
stars.

Using our Monte Carlo code we are also able to reproduce the intermediate
power-law slopes in the surface brightness profile seen in all $N$-body
simulations with IMBH in \citet{2005ApJ...620..238B}. Although, there
is considerable scatter in the profile which makes the determination
of the individual slopes rather uncertain, we  show that they, nevertheless,
are within the same region as in the $N$-body simulations. Using
twice as many stars in our simulations than were used in 
\citet{2005ApJ...620..238B}
we have a high enough resolution to obtain a reliable fit, which turns
out to be close to the average slope found by \citet{2005ApJ...620..238B}.
This, once again, confirms that, despite differences in the details
of the dynamics in the innermost cusp region, the overall cluster
structure and evolution is rather well reproduced. 

Finally, by carrying out a large parameter survey, we were able to model,
for the first time, the surface brightness profile of NGC 5694 in
\citet{2006AJ....132..447N} and constrain the maximum mass of a possible IMBH
in the cluster center to $<1000\, {\rm M_\odot}$, which, in this case,
corresponds to $M_{BH}/M_{c}\lesssim 0.2\%$.  Given the photometric errors in
the observed profile, both models with and without an IMBH can be constructed
to match observations.  However, considering the surface brightness slopes,
there is a slight preference for models with an IMBH, as clusters without IMBH
rarely had slopes as large as observed. This was in spite of the presence of a
significant population of binaries ($\approx 8\%$ in the end), which has
produced larger variations in the direct $N$-body simulations of
\citet{VesperiniTrenti} containing a much lower number of stars. Clearly,
taking into account the rather large uncertainties in the individual fitted
slopes, a more thorough statistical comparison of the surface brightness
profiles is required to better assess the likelihood of the presence of an IMBH
in NGC 5694. So far the evidence for an IMBH is only marginal. Moreover, the
existence of an IMBH in this particular cluster seems inconsistent with the
static models of \citet{2007MNRAS.381..103M}, with NGC 5694 having a
concentration that is too large for the measured inner surface brightness
slope. However, considering possible errors in the value for the concentration
quoted in the Harris catalog \citep{1996AJ....112.1487H} and the rather large
uncertainty in the inner surface brightness slope, we find that this is not yet
conclusive either.

In summary, we find that our Monte Carlo code, which now includes
the effects of a central IMBH on the stellar distribution, is able
to reproduce the overall structure and evolution of single-mass as
well as of more realistic, multi-mass cluster models reasonably well
compared to direct $N$-body simulations. There are differences in
the disruption rates and IMBH masses which are most likely related
to the wandering of the IMBH, which is not included in the Monte Carlo
code. However, these differences become almost negligible for realistic
cluster models as the disruption rates are generally very low for
these cases. In addition, since the IMBH wandering radius scales with
$M_{BH}/M_{*}$  as $r_{w}\sim\sqrt{M_{BH}/M_{*}}$  \citep{2002ApJ...572..371C}
the influence of the IMBH motion is reduced for more massive clusters
with a similar $M_{BH}/M_{c}$. One process that still plays a role
for more massive clusters and which is not included in our Monte Carlo
simulations are the close gravitational interactions with cusp stars
which we plan to implement in the near future.

\acknowledgements{We thank Holger Baumgardt for providing us with data as well
as initial conditions of the $N$-body simulations and helpful discussions. We
also thank Rainer Spurzem for very helpful comments and discussions that
finally lead to the derivation of $\dot{n}_{w}/\dot{n}_{lc}$.  This work was
supported by NSF Grant AST-0607498 at Northwestern University.  JMF
acknowledges support from Chandra Postdoctoral Fellowship Award PF7-80047.
Support for program HST-AR-11779 was provided by NASA through a grant from the
Space Telescope Science Institute, which is operated by the Association of
Universities for Research in Astronomy, Inc., under NASA contract NAS 5-26555.
This research was supported in part by the National Science Foundation under
Grant No. PHY05-51164. This work was completed while all authors were
participants of the program {}``Formation and Evolution of Globular Clusters''
at KITP in spring 2009.}

\bibliographystyle{/home/stefan/.TeX/bibtex/apj}

\end{document}